\documentclass[11pt]{article}

\usepackage{amsmath}

\numberwithin{equation}{section}

\def\a{\alpha}
\def\b{\beta}
\def\g{\gamma}

\def\d{\delta}
\def\e{\epsilon}

\def\k{\kappa}
\def\l{\lambda}
\def\m{\mu}

\def\o{\omega}
\def\q{\theta}
\def\th{\theta}

\def\s{\sigma}

\def\G{\Gamma}

\def\L{\Lambda}

\def\Si{\Sigma}

\def\ad{{\dot \alpha}}
\def\bd{{\dot \beta}}
\def\gd{{\dot \gamma}}
\def\dd{{\dot \delta}}

\def\D{{\cal D}}
\def\Db{{\bar{\cal D}}}
\def\qb{{\bar \theta}}
\def\sb{{\bar \sigma}}
\def\Qb{{\bar Q}}
\def\etab{{\bar \eta}}
\def\psib{{\bar \psi}}
\def\kb{{\bar \kappa}}

\def\p{\partial}
\def\Box{\p^2}

\newcommand\lcomp[1]{{#1}\big|_{\theta=\bar{\theta}=0}}

\newcommand{\ns}{\normalsize}

\begin{document}


\begin{titlepage}
\title{\hfill{\ns UPR-741-T, IASSNS-HEP-97/26, PUPT-1691\\[.1cm]}
   \hfill{\ns March 1997}\\[.8cm]
   {\LARGE Membranes and Three-form Supergravity}}
\author{Burt A.~Ovrut$^1\; ^2$ and Daniel Waldram$^3$\\[0.5cm]
   \emph{\ns $^1$ Department of Physics, University of Pennsylvania}\\[-1ex]
   \emph{\ns Philadelphia, PA 19104--6396, USA}\\[1ex]
   \emph{\ns $^2$ School of Natural Sciences, Institute for Advanced 
       Study}\\[-1ex]
   \emph{\ns Olden Lane, Princeton, NJ 08540, USA}\\[1ex]
   \emph{\ns $^3$ Department of Physics, Joseph Henry Laboratories}\\[-1ex]
   \emph{\ns Princeton University, Princeton, NJ 08544, USA }}
\date{}
\maketitle

\begin{abstract} 
\thispagestyle{empty}

We discuss membranes in four-dimensional $N=1$ superspace. The
$\kappa$-invariance of the Green-Schwarz action implies that there is
a dual version of $N=1$ supergravity with a three-form potential. We
formulate this new supergravity in terms of a three-form superfield in
curved superspace, giving the relevant constraints on the field
strength. We find the corresponding membrane soliton in the new
supergravity and discuss how the extended supersymmetry algebra
emerges from the symmetries of the flat superspace background. 

\end{abstract}

\thispagestyle{empty}

\end{titlepage}


\section{Introduction}


As was first pointed out some years ago \cite{AETW}, there is a nexus of 
relations connecting the existence of consistent Green-Schwarz super
$p$-brane actions and of supergravity theories containing $(p+1)$-form
potentials. The purpose of the present paper is to discuss the $p=2$
membrane and the corresponding supergravity in four dimensions. 
The membrane naturally couples to a three-form potential. We derive a
corresponding three-form version of $N=1$ supergravity in
superfields. It is a dualized version of old minimal supergravity
where one of the scalar auxiliary fields is replaced by the four-form
field strength of the three-form potential. The theory is a toy model
of the theory of a membrane in eleven dimensions, the three-form
supergravity sharing some of the structure of eleven-dimensional
supergravity. One advantage is that the four-dimensional theory
is an off-shell description, with all necessary auxiliary
fields. One notes that it is also strictly this three-form
version of $N=1$ supergravity that appears in the reduction of
higher-dimensional supergravities containing form potentials of degree
three or higher, as was discussed in \cite{ANT} in the context of $N=8$
supergravity. 

Consider starting with the theory of a fundamental
Green-Schwarz $p$-brane, which has manifest spacetime
supersymmetry. Consistent actions are $\kappa$-symmetric. That is,
they have a local fermionic world-volume symmetry which can be used to
gauge away half of the fermionic degrees of freedom. When such a
symmetry exists it is possible to go to a gauge where the action
exhibits world-volume supersymmetry. In this sense, the presence of
$\kappa$-symmetry allows the supersymmetry to be projected down from
the spacetime onto the world volume. 

The requirement of $\k$-symmetry puts strong constraints on the
structure of the target superspace. First, it implies that there is a
non-trivial background super $(p+1)$-form potential $B_{p+1}$
\cite{HM,HLP,BST}. The simplest action, proportional to the volume the
$p$-brane sweeps out as it moves through the target superspace, is not
$\kappa$-symmetric. A second, Wess-Zumino term, describing the
coupling of the $p$-brane to $B_{p+1}$, must be included. The field
strength $\Si=dB_{p+1}$ is required to satisfy a set of constraints
\cite{BST} so that, even in flat space, $B_{p+1}$ is not zero. These
constraints cannot be solved in a general $D$-dimensional
spacetime. One finds, for instance, that the membrane is only
consistent in $D=4$, $5$, $7$ and $11$ dimensions and always has $N=1$
target-space supersymmetry. In addition, the existence of a
$\k$-symmetry implies that the torsion, given in terms of the
supervielbein of the target superspace, is also constrained. 

Thus $\k$-symmetric $p$-brane actions exist only in superspace
backgrounds with a non-trivial super $(p+1)$-form. One might ask if
the constrained superform and the supervielbein of the
background superspace together form some supergravity multiplet. In
various cases, this has been shown to be true. For a membrane in
eleven dimensions, the constraints on $\Si$ and the torsions reproduce the
equations of motion of $N=1$, $D=11$ supergravity \cite{BST,DHIS}. The
five-brane in ten dimensions gives the constraints of $N=1$, $D=10$
supergravity formulated in terms of a six-form, the dual of the usual
anti-symmetric tensor \cite{AETW}. The the authors of \cite{AETW} also
point out that the three-form coupling to the membrane in four
dimensions can be considered as a special type of auxiliary field for
$D=4$, $N=1$ supergravity but do not further discuss the structure of
this non-standard off-shell supermultiplet. 

Alternatively one can start with a supergravity theory with a
$(p+1)$-form potential. As is now well-known, $p$-brane states then
appear as topological defects, solitons carrying either electric or
magnetic $(p+1)$-form charge (see for instance \cite{DKL}). When the
bosonic zero modes of the defect are all world-volume scalars, the 
fluctuations of the $p$-brane are described by the Green-Schwarz
action. 

An important property of the simplest $p$-brane solitons is that they
preserve one-half of the supersymmetries. As such, they are not
representations of the usual super-Poincar\'e algebra. Instead, they
are described by an algebra extended by a topological charge,
proportional to the $(p+1)$-form charge of the $p$-brane. Again this
extension can be seen either in terms of the supergravity solution
(see for instance \cite{AGITa,DGHR}) or as a topological charge in the
supersymmetry algebra of the Green-Schwarz action \cite{AGITb}.

The membrane we will consider in four dimensions is the analog of the
eight-brane in ten dimensions. In each case the $p$-brane is a domain wall
dividing the spacetime into two parts. The introduction of a nine-form
potential for the eight-brane \cite{8brane} required the dualization
of the massive form of supergravity due to Romans \cite{Romans}. In the
same way, we will find that the dualization of old minimal to
three-form supergravity naturally arises in the `massive' form of the
theory: that is, the action with a cosmological constant. In fact, the
supergravity solution corresponding to the four-dimensional membrane
is simply the patching of two pieces of anti-de Sitter space across a
domain wall. (Similar solutions with matter are familiar in the domain
wall literature \cite{domainwall}.) Domain-wall, $(D-2)$-branes were
discussed in six and seven dimensions in \cite{LPSS}, and then more 
generally in \cite{LPT,CLPST}. The latter papers include the solution
corresponding to a membrane in $D=4$, $N=1$ supergravity but do not
discuss what form of the corresponding dualized supergravity might
take.

The paper is arranged as follows. In section two we discuss the matter
three-form supermultiplet in flat superspace, first in components, as the
dualization of a chiral superfield, and then as a constrained super
three-form field, as was first given in \cite{Gates}. The
fields of the dualized chiral multiplet correspond to the full
three-form multiplet in a particular Wess-Zumino gauge. In section
three we give a self-contained derivation of the three-form version of
$N=1$ supergravity multiplet which
naturally couples to the membrane in four dimensions. We start with
old minimal supergravity multiplet in
components. The three-form theory can be obtained by simply dualizing
one of the scalar auxiliary fields in a way very similar to the
chiral-field dualization of section two. It is interesting to note
that a version of supergravity with both scalar auxiliary fields
replaced by three-forms was given in components in early studies of 
supergravity \cite{StelleWest,Ogiev}. A deeper understanding of the three-form 
theory and its gauge symmetries can only be
obtained in superspace. To this end, we then discuss the three-form
superfield $B$ coupled to old minimal supergravity in curved superspace,
following a recent paper by Bin\'etruy \textit{et al}
\cite{BPGG}. (These authors treat the three-form supermultiplet as
matter, rather than try to construct an irreducible supergravity multiplet
involving the three-form.) We present and solve the additional
constraints on $B$ required to merge the old minimal supergravity multiplet
with the matter three-form multiplet to produce irreducible off-shell
three-form supergravity. This is the three-form analog of the super 
two-form formalism used in \cite{newminimal} to construct new-minimal
supergravity. We discuss the general gauge structure of this theory
and Wess-Zumino gauge. The local supersymmetry algebra on
the component fields corresponds to the usual superdiffeomorphism
transformations together with a WZ-gauge-restoring transformation. 
The superdiffeomorphisms and gauge transformations are distinct 
symmetries, so that, in this superspace formulation, the three-form 
supergravity in no way corresponds to gauging the extended form of the
global superalgebra. The next two sections make the connection to
supermembranes in four dimensions. In section four we show that the
three-form theory naturally arises in dualizing the massive action of 
old minimal supergravity, namely the theory with a cosmological constant. We
briefly present the elementary membrane solution, match it to a
delta-function membrane source and show it preserves half
the supersymmetries. In section five we consider the fundamental
Green-Schwarz membrane action. The requirement of
$\k$-symmetry in four dimensions leads to exactly the super
three-form and torsion constraints we used in section three to
construct the three-form supergravity. Section six describes how
the extended supersymmetry algebra appears from the symmetries of a flat
supergravity background. We stress how the result depends on the
choice of background by also investigating the algebra in an anti-de
Sitter background. 

Throughout the paper we will use the conventions of Wess and Bagger
\cite{WessBagger} and the authors of \cite{BPGG}, except that we define
the Ricci tensor as $R_{mn}=\p_p{\G^p}_{mn}+\dots$.


\section{The three-form supermultiplet in flat superspace}


The supermembrane naturally couples to a three-form potential. We
start, in this section, therefore by simply describing the structure
of the three-form matter supermultiplet in flat space. It is most
easily derived in components by dualizing one of the auxiliary fields
of a chiral multiplet. The superspace formulation, on the other hand,
requires a full three-form superfield \cite{Gates,BPGG}

Consider a chiral superfield,
\begin{equation}
   \Phi=Y + \sqrt{2}\theta\eta + \theta^{2}F + \dots
\label{eq:2-1}
\end{equation}
The component fields $Y$, $\eta_{\alpha}$ and $F$ are known to
transform under supersymmetry as 
\begin{equation}
\begin{aligned}
   \delta_{\xi}Y &= \sqrt{2}\xi\eta \\
   \delta_{\xi}\eta &= i\sqrt{2}\sigma^{m}\bar{\xi}\partial_{m}Y
       + \sqrt{\xi}F \\
   \delta_{\xi}F &= i\sqrt{2}\bar{\xi}\bar{\sigma}^{m}\partial_{m}\eta 
\end{aligned}
\label{eq:2-2}
\end{equation}
We now want to write the auxiliary field $F$ as
$F=H+iF_{2}$, where both $H$ and $F_{2}$ are real fields. Then the
last equation in \eqref{eq:2-2} becomes
\begin{equation}
   \delta_{\xi}H = \frac{i}{\sqrt{2}} \left( \bar{\xi}\bar{\sigma}^{m}
      \partial_{m}\eta + \xi \s^m \p_m\bar{\eta} \right)
\label{eq:2-3}
\end{equation}
and
\begin{equation}
   \delta_{\xi}F_{2} = \frac{1}{\sqrt{2}}\left( \bar{\xi}\bar{\sigma}^{m}
      \partial_{m}\eta - \xi\s^m\p_m\bar{\eta} \right)
\label{eq:2-4}
\end{equation}
Now, we can always choose to express the real field $F_{2}$ as 
$\epsilon^{mnpq}F_{mnpq}$ where $F$ is a four-form. Since a four-form
in four dimensions is always closed, one can write locally $F=dC$
where $C$ is a three-form. That is, we can always write 
\begin{equation}
   F_{2} = -\frac{4}{3}\epsilon^{mnpq}\partial_{m}C_{npq}
\label{eq:2-5}
\end{equation}
where we have chosen the normalization $-4/3$ to conform to the
conventions of reference \cite{BPGG}. Inserting this expression in 
\eqref{eq:2-4}, we find that this equation can be completely integrated. 
The result is that
\begin{equation}
   \delta C_{mnp} = -\frac{\sqrt{2}}{16}\epsilon_{mnpq}
      \left( \bar{\xi}\bar{\sigma}^{q}\eta - \xi\sigma^{q}\bar{\eta} \right)
\label{eq:2-6}
\end{equation}
Putting everything together we have 
\begin{equation}
\begin{aligned}
   \delta Y &= \sqrt{2}\xi\eta \\
   \delta\eta &= i\sqrt{2}\sigma^{m}\bar{\xi}\partial_{m}Y
      + \sqrt{2} \xi F \\
   \delta H &= \frac{i}{\sqrt{2}} 
      \left( \bar{\xi}\bar{\sigma}^{m}\partial_{m}\eta - 
      \partial_{m}\bar{\eta}\bar{\sigma}^{m}\xi \right) \\
   \delta C_{mnp} &= -\frac{\sqrt{2}}{16}\epsilon_{mnpq}
      \left( \bar{\xi}\bar{\sigma}^{q}\eta - \xi\sigma^{q}\bar{\eta} \right)
\end{aligned}
\label{eq:2-7}
\end{equation}
Note that we have dropped the subscript $\xi$ in writing these equations. The
reason for this will become clear below. It follows that the component fields
$Y$, $\eta_{\alpha}$, $H$ and $C_{mnp}$ form an irreducible
supermultiplet, a three-form matter multiplet. One
could, of course, have dualized the other auxiliary field $H$, or
some linear combination of $H$ and $F_2$. However, as we will see, all
these dualizations are simply reparameterizations of the same
underlying multiplet. Dualizing both auxiliary field does lead to a
completely new multiplet \cite{OneandOne}. However it is not
relevant to the coupling of the membrane to supergravity. 

What is the superfield for which $Y$, $\eta_{\alpha}$, $H$ and
$C_{mnp}$ are the component fields? To
answer this, we must consider the theory of a constrained three-form
supermultiplet \cite{Gates}. We will use the notation and results from
reference \cite{BPGG}. One begins by introducing a real three-form
superfield 
\begin{equation}
   B = \frac{1}{3!}e^{A}e^{B}e^{C}B_{CBA}
\label{eq:2-8}
\end{equation}
where $e^{A}$ denotes the frame of flat superspace. The invariant field
strength is a four-form 
\begin{equation}
   \Sigma = dB
\label{eq:2-9}
\end{equation}
Its component superfields, defined by
\begin{equation}
   \Sigma = \frac{1}{4!}e^{A}e^{B}e^{C}e^{D}\Sigma_{DCBA}
\label{eq:2-10}
\end{equation}
are then subject to the constraints
\begin{equation}
   \Sigma_{\underline{\d\g\b}A} = 0
\label{eq:2-11}
\end{equation}
where $\underline{\alpha}\sim\alpha$, $\dot{\alpha}$, together with
the conventional constraint
\begin{equation}
   {\Si^\dd}_{\g ba} = 0
\end{equation}
The Bianchi identity $d\Sigma=0$ can be solved subject to these
constraints. One finds that all the coefficients of the four-form
$\Sigma$ either vanish, or can be expressed in terms of a complex
superfield $Y$ and and its conjugate $\bar{Y}$ as follows 
\begin{align}
   \Sigma_{\d\g ba} &= 
      \frac{1}{2}\left(\s_{ba}\e\right)_{\d\g}\bar{Y} \; , \label{eq:2-12} \\
   {\Sigma^{\dd\gd}}_{ba} &= 
      \frac{1}{2}\left(\bar{\s}_{ba}\e\right)^{\dd\gd}Y \label{eq:2-13}
\end{align}
and
\begin{align}
   \Sigma_{\d cba} &= 
      - \frac{1}{16}{\s^d}_{\d\dd}\e_{dcba}\bar{D}^{\dd}\bar{Y} 
      \; , \label{eq:2-14} \\
   {\Sigma^\dd}_{cba} &= 
      \frac{1}{16}{\sb}^{d \dd\d}\e_{dcba}D_{\d}Y \label{eq:2-15}
\end{align}
while
\begin{equation}
   \Si_{dcba} = \frac{1}{12}i\e_{dcba} \left(D^{2}Y-\bar{D}^{2}\bar{Y}\right)
\label{eq:2-17}
\end{equation}
In addition, the superfield $Y$ and it conjugate $\bar{Y}$ are subject
to the chirality conditions
\begin{equation}
   \bar{D}^{\dot{\alpha}}Y = 0 \; , \qquad  D_\a \bar{Y} = 0 
\label{eq:2-16}
\end{equation}

One can also solve the constraints \eqref{eq:2-11} directly to find the
coefficients of the super three-form in terms of a real, but otherwise 
unconstrained, prepotential, $\Omega$. The result is that there is a
gauge where all coefficients vanish with the exception of
\begin{gather}
   B_{\g\ \;\ a}^{\ \;\bd} = 
      -2i{\left(\s_{a}\e\right)_\g}^{\bd}\Omega \label{eq:2-18} \\
   B_{\g ba} = 
      2{\left(\s_{ba}\right)_\g}^\b D_\b \Omega \; , \quad
   {B^\gd}_{ba} = 
      2{\left(\bar{\s}_{ba}\right)^\gd}_\bd \bar{D}^\bd \Omega \label{eq:2-19}
\end{gather}
and
\begin{equation}
   B_{cba} = \frac{3}{8}\e_{dcba}\bar{\s}^{d\ad\a} 
      \left[D_\a,\bar{D}_\ad\right] \Omega
\label{eq:2-20}
\end{equation}
Substituting these expressions back into the field strength yields the
expressions
\begin{align}
   \bar{Y} &= -4D^{2}\Omega \label{eq:2-21} \\
   Y &= -4\bar{D}^{2}\Omega \label{eq:2-22}
\end{align}
reflecting the fact that $Y$ is chiral. 
It is important to note that there is a residual gauge freedom in
$B$. The fields $Y$ and $\bar{Y}$ remain unchanged under the transformation
\begin{equation}
   \delta_{L}\Omega=L
\label{eq:2-23}
\end{equation}
where $L$ is a real linear multiplet satisfying the constraints
\begin{equation}
   D^{2}L = \bar{D}^{2}L = 0
\label{eq:2-24}
\end{equation}
If we now define
\begin{equation}
   \lcomp{B_{cba}} = C_{cba}
\label{eq:2-25}
\end{equation}
as well as
\begin{equation}
\begin{gathered}
   \lcomp{Y} = Y \; , \quad \lcomp{\bar{Y}} = \bar{Y} \\
   \lcomp{D_{\alpha}Y} = \sqrt{2}\eta_{\alpha}\; , \quad
   \lcomp{\bar{D}_\ad Y} = \sqrt{2}\bar{\eta}_{\ad}
\label{eq:2-26}
\end{gathered}
\end{equation}
and
\begin{equation}
   \lcomp{D^2 Y} + \lcomp{\bar{D}^2 \bar{Y}} = - 8H
\label{eq:2-27}
\end{equation}
then, using \eqref{eq:2-17}, we can identify $Y$ with the dualized
chiral superfield $\Phi$ where 
\begin{equation}
   F = H - i\frac{4}{3}\epsilon^{mnpq}\partial_{m}C_{npq}
\label{eq:2-28}
\end{equation}
We recall that we could have dualized any linear combination of the
auxiliary fields in $\Phi$. We now see that the resulting multiplet
corresponds to a parameterization where we identify $Y$ with
$e^{i\a}\Phi$ for some constant phase $\a$. Henceforth, we 
will assume that the simple identification $Y=\Phi$ is made.
In this case, the general structure of an unconstrained real superfield 
$\Omega$ satisfying equations \eqref{eq:2-21} and \eqref{eq:2-22} is given by
\begin{equation}
\begin{split}
   \Omega = C + i\theta\chi - i\bar{\theta}\bar{\chi} 
      + \frac{1}{16}\theta^{2}Y^{*} + \frac{1}{16}\bar{\theta}^{2}Y
      + \frac{1}{6}\theta\sigma^{m}\bar{\theta}\epsilon_{mnpq}C^{npq}
      + \frac{1}{2}\theta^{2}\bar{\theta} 
          \left( \frac{\sqrt{2}}{8}\bar{\eta} + \bar{\s}^{m}\p_{m}\chi \right) 
      \\
      + \frac{1}{2}\bar{\theta}^{2}\theta
          \left( \frac{\sqrt{2}}{8}\bar{\eta} - \s^{m}\p_{m}\bar{\chi} \right) 
      + \frac{1}{4} \theta^{2} \bar{\theta}^{2} 
          \left(\frac{1}{4}H -\Box C \right)
\end{split}
\label{eq:2-29}
\end{equation}
We see that, in addition to $Y$, $\eta_{\alpha}$, $H$ and $C_{mnp}$, the
prepotential generically contains the two extra fields $C$ and $\chi$.
However, by an appropriate choice of $L$ these last two fields can be
transformed away to give
\begin{equation}
   \Omega = \frac{1}{16}\theta^{2}Y^{*} + \frac{1}{16}\bar{\theta}^{2}Y
      + \frac{1}{6}\theta\sigma^{m}\bar{\theta}\e_{mnpq}C^{npq} 
      + \frac{1}{8\sqrt{2}}\theta^{2}\bar{\theta}\bar{\eta}
      + \frac{1}{8\sqrt{2}}\bar{\theta}^{2}\theta\bar{\eta}
      + \frac{1}{16}\theta^{2}\bar{\theta}^{2}H 
\label{eq:2-30}
\end{equation}
This is the WZ gauge form for the superfield $\Omega$. 

We can now answer our original question. The 
fields $Y$, $\eta_{\alpha}$, $H$ and $C_{mnp}$ are the
component fields of the superfield $\Omega$ in WZ gauge. Let us now 
consider the supersymmetry transformation of this superfield. It is a
simple exercise to show that $\delta_{\xi}\Omega$ contains terms
proportional to $\theta$ and $\bar{\theta}$ and, hence, is no longer in WZ
gauge. However, WZ gauge can be restored by a gauge transformation
$\delta_{L}\Omega$ where the component fields $B$, $\lambda_{\alpha}$ and
$B_{mn}$ of $L$ are chosen to satisfy
\begin{equation}
\begin{gathered}
   B = 0 \\
   \lambda = \frac{1}{6}i\sigma^{m}\bar{\xi}\e_{mnpq}C^{npq} 
       + \frac{i}{8} \xi Y^{*} \\
   \p_{[m}B_{np]} = 0
\end{gathered}
\label{eq:2-31}
\end{equation}
It is not hard to verify that, under supersymmetry plus the WZ 
gauge-restoring transformation, the component fields transform as 
\begin{equation}
\begin{aligned}
   \left(\delta_{\xi}+\delta_{L}\right) Y &= \sqrt{2}\xi\eta \\
   \left(\delta_{\xi}+\delta_{L}\right) \eta &= 
       i\sqrt{2}\sigma^{m}\bar{\xi}\partial_{m}Y + \sqrt{2} \xi F \\
   \left(\delta_{\xi}+\delta_{L}\right) H &= 
       \frac{i}{\sqrt{2}} \left( \bar{\xi}\bar{\s}^m\p_m\eta 
           + \xi\s^m\p_m\bar{\eta} \right) \\
   \left(\delta_{\xi}+\delta_{L}\right) C_{mnp} &= 
       - \frac{\sqrt{2}}{16}\e_{mnpq}
           \left( \bar{\xi}\bar{\s}^q \eta - \xi \s^q \bar{\eta} \right)
\end{aligned}
\label{eq:2-32}
\end{equation}
Comparing these results with equation \eqref{eq:2-7}, we see that
\begin{equation}
   \delta = \delta_{\xi} + \delta_{L}
\label{eq:2-33}
\end{equation}
This explains why we dropped the subscript $\xi$ in \eqref{eq:2-7}. 

We conclude that the supermultiplet obtained by integrating the supersymmetry
transformations of the component fields of $\Phi$ is precisely the
vector prepotential $\Omega$ in the WZ gauge. Furthermore, the transformation
law of the component fields after this integration correspond to
the combined supersymmetry and WZ gauge-restoring transformations of the
prepotential supermultiplet.
It is useful to note, using \eqref{eq:2-18}--\eqref{eq:2-20}, that for 
$\Omega$ in WZ gauge, the lowest components of the superfield coefficients 
$B_{CBA}$ satisfy
\begin{gather}
   \lcomp{B^{\ \;\bd}_{\g\ \ \; a}} = 
   \lcomp{B_{\g b a}} =
   \lcomp{{B^\gd}_{ba}} = 0 \label{eq:2-34} \\
   \lcomp{B_{cba}} = C_{cba} \label{eq:2-35}
\end{gather}
If $\Omega$ is transformed outside of its WZ gauge, non-zero components begin
to appear in \eqref{eq:2-34}. Thus, WZ gauge for the prepotential corresponds
to the gauge one would naturally define for the coefficients $B_{CBA}$. 

We conclude this section by computing the commutator of the transformations 
$\delta$ acting on the component field $C_{mnp}$. We find that
\begin{equation}
   \left(\d \d'-\d' \d \right) C_{mnp} = 
       2\eta^{q}\partial_{q}C_{mnp} + 3\p_{[m}\L_{np]}
\label{eq:2-36}
\end{equation}
where 
\begin{equation}
   \Lambda_{mn} = -2C_{mnp}\eta^{p} 
       - \frac{1}{2} \left( \eta^{+}_{pq}Y_{1} + \eta^{-}_{pq}Y_{2} \right)
\label{eq:2-42}
\end{equation}
with
\begin{equation}
\begin{aligned}
   \eta^{m} &= -i \left( \xi\s^m\bar{\xi}' - \xi'\s^m\bar{\xi} \right) \\
   \eta^{+}_{pq} &= \xi\s_{pq}\xi' + \bar{\xi}\bar{\s}_{pq}\bar{\xi}' \\
   \eta^{-}_{pq} &= -i \left( \xi\s_{pq}\xi' 
       - \bar{\xi}\bar{\s}_{pq}\bar{\xi}' \right) \\
\end{aligned}
\label{eq:2-37}
\end{equation}
and $Y=Y_{1}+iY_{2}$. What is the interpretation of the two terms on the right
hand side of equation \eqref{eq:2-36}? Substituting
$\delta=\delta_{\xi}+\delta_{L}$ on the left side of equation \eqref{eq:2-36},
and using the fact that $\d_L\d_{L'}C_{mnp}=\d_L\d_{\xi}C_{mnp}=0$, we find
\begin{equation}
   \left( \d_\xi\d_{\xi'} - \d_{\xi'}\d_\xi \right) C_{mnp} = 
      2\eta^{q}\partial_{q}C_{mnp}
\label{eq:2-38}
\end{equation}
and
\begin{equation}
   \left( \d_\xi\delta_{L'} - \d_{\xi'}\d_L \right) C_{mnp} = 
      3\p_{[m}\L_{np]}
\label{eq:2-39}
\end{equation}
The first of these equations is simply the representation of the supersymmetry
algebra $\left\{Q_{\alpha},Q_{\dot{\alpha}}\right\}=
2{\s^m}_{\alpha\dot{\alpha}}P_{m}$
acting on the component field $C_{mnp}$. Hence, the first term on the right
side of equation \eqref{eq:2-36} is the translation part of the superalgebra.
If WZ gauge restoration was unnecessary, then this would be the only term to
appear. However, WZ gauge restoration is necessary and \eqref{eq:2-39}
tells us that the effect of this gauge restoration is the second term on the
right hand side of \eqref{eq:2-36}. The main point is this. One might
try to interpret the second term in  \eqref{eq:2-36} as indicating
the presence of central charges in the supersymmetry algebra associated with
the supermultiplet $\Omega$. These central charges would be Lorentz indexed,
$Z_{mn}$, and would extend the supersymmetry algebra to
\begin{equation}
   \left\{ Q_{\alpha}, Q_{\beta} \right\} = 
       2i\left(\sigma^{mn}\e\right)_{\alpha\beta} Z_{mn}
\label{eq:2-40}
\end{equation}
The second term in \eqref{eq:2-36} would then be viewed as a (field-dependent)
gauge transformation
\begin{equation}
   \delta_{gauge}C_{mnp} = 3\partial_{[m}\Lambda_{np]}
\label{eq:2-41}
\end{equation}
associated with these extended charges. One problem is that the
algebra does not then close simply. However, it is clear from the
above analysis that this interpretation is unwarranted, that the
extra term is an artifact of the WZ gauge and is perfectly consistent
with the usual, unextended supersymmetry algebra.


\section{$N=1$ three-form supergravity}


In this section we give a self-contained description of the three-form
version of $N=1$ supergravity which naturally couples to the membrane
in four dimensions. We first give the multiplet in components, by
dualizing one of the auxiliary fields of the old minimal supergravity
multiplet. We then turn to the full description in terms of
superfields. 

A geometrical construction of any $N=1$ supergravity multiplet starts
by introducing a supervielbein, ${E_M}^A$, and a superconnection
one-from ${\phi_A}^B$ to gauge local Lorentz transformations
\cite{WessBagger}. One then forms the superfield curvature,
${R_A}^B=d{\phi_A}^B+{\phi_A}^C\wedge{\phi_C}^B$, 
and the supertorsion, $T^{A}=dE^{A} + E^{B}\wedge{\phi_B}^A$. The usual,
so-called old minimal, form of the supergravity multiplet is obtained by
constraining particular components of the torsion superfield. These
constraints are summarized in equation \eqref{Tconstraints} later in
this section. One then solves the Bianchi identities for ${R_A}^B$ and
$T^A$ subject to these constraints. The result is
that all superfield components of the curvature and torsion that do not vanish
can be constructed from three fundamental superfields, $R$, $G_{\alpha
\dot{\alpha}}$ and $W_{\alpha\beta\gamma}$. The component fields of the old
minimal gravity supermultiplet are obtained as follows. First,
one uses part of the local Lorentz and diffeomorphic symmetries to
rotate the lowest component of the supervielbein into the form
\begin{equation}
   \lcomp{{E_M}^{A}} = \left( \begin{array}{ccc}
       {e_m}^{a}   & \frac{1}{2}{\psi_m}^\a 
            & \frac{1}{2}\bar{\psi}_{m \ad} \\
       0   & {\delta_\mu}^\a  & 0 \\
       0   & 0   & {\delta^{\dot{\mu}}}_\ad 
       \end{array} \right)
\label{eq:3-1}
\end{equation}
A vielbein in this form is said to be in WZ gauge. One also defines a
complex scalar field 
\begin{equation}
   \lcomp{R} = -\frac{1}{6}M
\label{eq:3-2}
\end{equation}
and a real vector field
\begin{equation}
   \lcomp{G_{a}} = -\frac{1}{3}b_{a}
\label{eq:3-3}
\end{equation}
The component fields of the old minimal supergravity multiplet are then
${e_m}^a$, ${\psi_m}^\a$, $b_{a}$ and $M$. Generic fermionic
superdiffeomorphisms, $\delta_{\xi}$, parameterized by a supervector
field $\xi^\a$, take one outside of the WZ
gauge. However, the gauge can be re-established using a restoring 
transformation, $\delta_{WZ}$, as discussed, for example, in 
\cite{WessBagger}. It is customary to define, 
$\delta=\delta_{\xi}+\delta_{WZ}$, and to refer to these
variations as supersymmetry transformations. The variation of the
component fields under supersymmetry are 
\begin{equation}
\begin{aligned}
   \delta {e_m}^a &= i \left( \psi_{m}\sigma^{a}\bar{\xi} 
       + \bar{\psi}_m\sb^{a}\xi \right) \\
   \delta {\psi_m}^\a &= -2\D_m\xi^\a  
       + i{e_m}^{c} \left\{ 
           \frac{1}{3} M \left(\e\s_c\bar{\xi}\right)^\a 
           + b_{c}\xi^\a
           + \frac{1}{3} b^{d} \left(\xi\s_d\bar{\s}_{c}\right)^\a 
           \right\} \\
   \delta b_{\alpha\dot{\alpha}} &= 
       \xi^{\delta} \left\{
           \frac{3}{4} \bar{\psi}_{\a\ \;\d\gd\ad}^{\ \gd}
           + \frac{1}{4} \e_{\d\a} \bar{\psi}^{\g\gd}_{\ \ \;\g\ad\gd}
           - \frac{i}{2} M^* \psi_{\a\ad\d}
           + \frac{i}{4} \left( 
               \bar{\psi}_{\a\dot{\rho}}^{\ \ \;\dot{\rho}} b_{\d\ad} 
               + \bar{\psi}_{\d\dot{\rho}}^{\ \ \;\dot{\rho}} b_{\a\ad} 
               - \bar{\psi}_{\d\ \;\ad}^{\ \dot{\rho}} b_{\a\dot{\rho}} \right)
           \right\} + \text{h.c.} \\ 
   \delta M &= - \xi \left( \s^{a}\bar{\s}^{b}\psi_{ab} 
       + ib^{a}\psi_{a} -i\s^a\bar{\psi}_a M \right) 
\end{aligned}
\label{eq:3-4}
\end{equation}
We have included the variation of $b_a$ for completeness, though it is
not required in the following. The spinor notation used for $\d b_a$ is
explained in \cite{WessBagger}. 
Writing the auxiliary field $M$ as $M=M_{1}+iM_{2}$, the last equation in
\eqref{eq:3-4} becomes
\begin{equation}
\begin{split}
   \delta M_{1} &= - \xi\s^{ab}\psi_{ab} - \bar{\xi}\sb^{ab}\bar{\psi}_{ab}
       - \frac{i}{2} \left( \xi\psi_a - \bar{\xi}\bar{\psi}_a \right) b^a
       \\ & \quad
       + \frac{i}{2} \left( \xi\s^a\bar{\psi}_a 
           + \bar{\xi}\sb^a\psi_a \right) M_{1}
       - \frac{1}{2} \left( \xi \s^a \bar{\psi}_a 
           - \bar{\xi}\sb^a\psi_a \right) M_{2}
\end{split}
\label{eq:3-5}
\end{equation}
and
\begin{equation}
\begin{split}
   \delta M_{2} &= i \left( \xi\s^{ab}\psi_{ab} 
           - \bar{\xi}\sb^{ab}\bar{\psi}_{ab} \right)
       - \frac{1}{2} \left( \xi\psi_a + \bar{\xi}\bar{\psi}_a \right) b^a
       \\ &
       + \frac{1}{2} \left( \xi \s^a \bar{\psi}_a 
           - \bar{\xi}\sb^a\psi_a \right) M_{1}
       + \frac{i}{2} \left( \xi \s^a \bar{\psi}_a
           + \bar{\xi}\sb^a\psi_a \right) M_{2}
\end{split}
\label{eq:3-6}
\end{equation}

In analogy with the previous section, we will try to introduce a
three-form field into the multiplet by dualizing the real field
$M_{2}$. We first write $M_{2}\propto\epsilon^{mnpq}F_{mnpq}$ where
$F$ is a four-form. Since a four-form in four-dimensions is always
closed, one has locally $F=dC$ where $C$ is a three-form. We can then
hope to integrate the supersymmetry transformation to give the
variation of $C$. Here, however, the exact analogy with the flat
superspace chiral multiplet case comes to an end. It turns out that,
in the supergravity case, it is necessary to add a second term to $dC$
in order to insure integrability. This term naturally arises when one
tries to dualize the old minimal supergravity action with a
cosmological constant, as we shall see in the next section. The
appropriate expression is
\begin{equation}
   M_{2} = \e^{mnpq}\partial_{m}C_{npq}
       + \frac{i}{2} \left( \psi_{a}\s^{ab}\psi_{b} 
           - \bar{\psi}_{a}\bar{\s}^{ab}\bar{\psi}_{b} \right)
\label{eq:3-7}
\end{equation}
where we have chosen unit normalization of the first term on the right hand
side for convenience. Inserting this in \eqref{eq:3-6}, we find that
the expression can be completely integrated. The result is that
\begin{equation}
   \delta C_{mnp} = \frac{i}{3}\e_{mnpq} \left( 
       \xi\s^{qr}\psi_{r} - \bar{\xi}\bar{\s}^{qr}\bar{\psi}_{r} \right)
\label{eq:3-8}
\end{equation}
This expression was actually first given by
Stelle and West in the one of the original formulations of off-shell $N=1$
supergravity \cite{StelleWest}. However in that case both the
components of $M$ were dualized. 
The proof that equation \eqref{eq:3-6} can be integrated is considerably more
involved than in the flat superspace case. In particular, there are several
of terms cubic in the gravitino field that arise which, 
at first sight, seem to
obstruct the integrability. However, by prodigious use of Fierz re-arrangement,
one can show that these terms all cancel. Putting everything together, we find
the variations

\begin{equation}
\begin{aligned}
   \delta {e_m}^a &= i \left( \psi_{m}\sigma^{a}\bar{\xi} 
       + \bar{\psi}_m\sb^{a}\xi \right) \\
   \delta {\psi_m}^\a &= -2\D_m\xi^\a  
       + i{e_m}^{c} \left\{ 
           \frac{1}{3}M \left(\e\s_c\bar{\xi}\right)^\a 
           + b_{c}\xi^\a
           + \frac{1}{3}b^{d} \left(\xi\s_d\bar{\s}_{c}\right)^\a 
           \right\} \\
   \delta b_{\alpha\dot{\alpha}} &= 
       \xi^{\delta} \left\{
           \frac{3}{4} \bar{\psi}_{\a\ \;\d\gd\ad}^{\ \gd}
           + \frac{1}{4} \e_{\d\a} \bar{\psi}^{\g\gd}_{\ \ \;\g\ad\gd}
           - \frac{i}{2} M^* \psi_{\a\ad\d}
           + \frac{i}{4} \left( 
               \bar{\psi}_{\a\dot{\rho}}^{\ \ \;\dot{\rho}} b_{\d\ad} 
               + \bar{\psi}_{\d\dot{\rho}}^{\ \ \;\dot{\rho}} b_{\a\ad} 
               - \bar{\psi}_{\d\ \;\ad}^{\ \dot{\rho}} b_{\a\dot{\rho}} \right)
           \right\} + \text{h.c.} \\ 
   \delta M_{1} &= - \xi\s^{ab}\psi_{ab} - \bar{\xi}\sb^{ab}\bar{\psi}_{ab}
       - \frac{i}{2} \left( \xi\psi_a - \bar{\xi}\bar{\psi}_a \right) b^a
       \\ & \quad
       + \frac{i}{2} \left( \xi\s^a\bar{\psi}_a 
           + \bar{\xi}\sb^a\psi_a \right) M_{1}
       - \frac{1}{2} \left( \xi \s^a \bar{\psi}_a 
           - \bar{\xi}\sb^a\psi_a \right) M_{2} \\
   \delta C_{mnp} &= \frac{i}{3}\e_{mnpq} \left( 
       \xi\s^{qr}\psi_{r} - \bar{\xi}\bar{\s}^{qr}\bar{\psi}_{r} \right)
\end{aligned}
\label{eq:3-9}
\end{equation}
It follows that the component fields
${e_m}^a$, ${\psi_m}^\a$, $b_{a}$, $M_{1}$ and $C_{mnp}$ 
form an irreducible supergravity multiplet. 

This fact is somewhat remarkable and 
deserves further discussion. As stated above, 
the old-minimal form of the $N=1$ supergravity multiplet consists of the 
component fields ${e_m}^a$, ${\psi_m}^\a$, $b_{a}$ and $M$. 
It has long been known that there exists another
irreducible $N=1$ supergravity multiplet, the so-called new minimal
supermultiplet, consisting of the component fields
${e_m}^a$ and ${\psi_m}^\a$ along with a real vector field
$a_{m}$ and a real two-form $B_{mn}$. That is, the auxiliary fields $M$ and
$b_{a}$ of the old minimal multiplet are replaced by auxiliary fields $a_{m}$
and $B_{mn}$ in the new minimal multiplet. Both the new and old minimal forms
of supergravity have 12 bosonic and 12 fermionic off-shell degrees of
freedom. There is one other irreducible $N=1$ supergravity multiplet known,
the so-called 16-16 multiplet, which, as the name implies, has 16 bosonic
and 16 fermionic off-shell degrees of freedom. It also contains 
a real two-form. However, none of these known multiplets contains a real
three-form. We seem, therefore, by writing $M_{2}$ as the curl of a three-form
and by integrating transformation equation \eqref{eq:3-6}, to have constructed
a new irreducible $N=1$ supergravity multiplet containing a three-form
$C_{mnp}$. This multiplet also has 12 bosonic and 12 fermionic off-shell
degrees of freedom. As we have mentioned, a component form of the supergravity 
multiplet with two three-form fields was derived some time ago in 
\cite{StelleWest,Ogiev}. The case of a single three-form multiplet is 
easily extracted from these papers, but was not explicitly discussed. 
This multiplet is interesting in that it is similar in
form, albeit in four dimensions, to eleven-dimensional supergravity. Unlike
eleven-dimensional supergravity, however, we can construct the 
off-shell theory and completely study its behaviour. In particular, and this 
is our main point, one can ask for what superfields are ${e_m}^a$, 
${\psi_m}^\a$, $b_{a}$, $M_{1}$ and $C_{npq}$ the component fields? 

To answer this, we must consider the theory of a
constrained three-form supermultiplet coupled to supergravity in curved
superspace. This construct is the analog of the super two-form
formalism used to \cite{newminimal} to derive the new minimal multiplet. 
The supergravity torsion constraints, and consequently the solution of
the geometrical Bianchi identities, remain the same as above. It follows that
the component fields arising from the geometrical supergravity sector
are just those defined in equations \eqref{eq:3-1}--\eqref{eq:3-3}.
Similarly, the transformations of these component fields under supersymmetry
are still given by \eqref{eq:3-4}--\eqref{eq:3-6}.
The theory of the super three-form, however, requires some modification 
over the flat
superspace case discussed in the previous section. This theory was provided in
detail in reference \cite{BPGG}, and we will use their notation and
some of their results in our
construction. It is worth emphasizing that these authors treated the 
three-form as a matter supermultiplet so did not use their formalism
to construct the new three-form irreducible gravity supermultiplet
found above. That construction requires imposing further constraints
on the three-form multiplet, as we will now describe.

Construction of the three-form matter multiplet in curved superspace
follows closely the flat-space case given in section two. One begins
by introducing a real three-form superfield
\begin{equation}
   B = \frac{1}{3!}E^{A}E^{B}E^{C}B_{CBA}
\label{eq:3-10}
\end{equation}
where now $E^{A}$ denotes the frame of curved superspace. The invariant field
strength is a four-form 
\begin{equation}
   \Sigma = dB = \frac{1}{4!}E^{A}E^{B}E^{C}E^{D}\Sigma_{DCBA}
\label{eq:3-11}
\end{equation}
Again the components of $\Si$ are constrained to satisfy
\begin{equation}
   \Sigma_{\underline{\d\g\b} A} = {\Si^\dd}_{\g ba} = 0
\label{eq:3-13}
\end{equation}
where $\underline{\alpha}\sim\alpha$, $\dot{\alpha}$. Solving the
Bianchi identity $d\Sigma=0$ one finds that all coefficients of the
four-form $\Sigma$ either vanish, or can be expressed in terms of an
complex superfields $Y$ and and its conjugate $\bar{Y}$ as follows 
\begin{align}
   \Sigma_{\d\g ba} &= 
      \frac{1}{2}\left(\s_{ba}\e\right)_{\d\g} \bar{Y} \; , \label{eq:3-14} \\
   {\Sigma^{\dd\gd}}_{ba} &= 
      \frac{1}{2} \left( \bar{\s}_{ba}\e\right)^{\dd\gd} Y \label{eq:3-15} 
\end{align}
and
\begin{align}
   \Sigma_{\d cba} &= -\frac{1}{16}
      {\s^d}_{\d\dd}\e_{dcba}\Db^{\dd} \bar{Y} \; , \label{eq:3-16} \\
   {\Sigma^\dd}_{cba} &= \frac{1}{16}
      \bar{\s}^{d \dd\d}\epsilon_{dcba}\D_{\d} Y \label{eq:3-17}
\end{align}
while 
\begin{equation}
   \Sigma_{dcba} = \frac{i}{12}\e_{dcba} \left[
      \left( {\cal{D}}^{2} - 24R^{\dagger} \right) Y
      - \left( \bar{\cal{D}}^{2}- 24R \right) \bar{Y} \right]
\label{eq:3-19}
\end{equation}
Furthermore, superfields $Y$ and $\bar{Y}$ are subject to the
curved-space chirality conditions
\begin{equation}
   \Db^{\dot{\alpha}}Y = 0 \; , \quad \D_{\alpha}\bar{Y} = 0
\label{eq:3-18}
\end{equation}

Also as before, one can solve the constraints \eqref{eq:3-13} directly
to find the coefficients of the super three-form in terms of a real
but otherwise  unconstrained prepotential, $\Omega$. One finds that there is a gauge where all coefficients vanish with the exception of
\begin{equation}
   B_{\g\ \; a}^{\ \bd} = -2i {\left(\sigma_{a}\e\right)_\g}^\bd \Omega
\label{eq:3-20}
\end{equation}
as well as
\begin{equation}
   B_{\g ba} = 
       2{\left(\s_{ba}\right)_\g}^\b D_{\beta}\Omega \; , \quad 
   {B^\gd}_{ba} = 
       2{\left(\bar{\s}_{ba}\right)^\gd}_\bd \Db^{\bd}\Omega
\label{eq:3-21}
\end{equation}
and
\begin{equation}
   B_{cba} = \frac{1}{4}\e_{dcba}\bar{\s}^{d\ad\a} \left(
       \left[ \D_\a, \Db_\ad \right] - 4G_{\a\ad} \right) \Omega
\label{eq:3-22}
\end{equation}
Substituting these expressions back into the field strength yields the
expressions
\begin{align}
   \bar{Y} &= -4 \left( {\cal{D}}^{2} - 8R^{\dagger} \right) \Omega 
      \label{eq:3-23} \\
   Y &= -4 \left( \bar{\cal{D}}^{2} - 8R \right) \Omega
      \label{eq:3-24}
\end{align}
The residual gauge freedom is reflected in the fact that the fields
$Y$ and $\bar{Y}$ remain unchanged under the transformation
\begin{equation}
   \delta_{L}\Omega=L
\label{eq:3-25}
\end{equation}
where $L$ is a real linear multiplet satisfying the constraints
\begin{equation}
   \left( {\cal{D}}^{2} - 8R^{\dagger} \right) L =
   \left( \bar{\cal{D}}^{2} - 8R \right)L = 0
\label{eq:3-26}
\end{equation}

The discussion of the WZ gauge for the three-form supermultiplet in curved
superspace is much the same as it was in the flat superspace case. The
prepotential $\Omega$
can be put into WZ gauge using transformation \eqref{eq:3-25} with an
appropriate choice of linear superfield $L$. In this gauge the lowest two
components of $\Omega$ vanish. It follows from 
\eqref{eq:3-20}--\eqref{eq:3-22} that in WZ gauge
\begin{gather}
   \lcomp{B^{\ \dot{\beta}}_{\alpha,a}} =
      \lcomp{B_{\alpha b a}} =
      \lcomp{B^{\dot{\alpha}}_{\ ba}} = 0 \label{eq:3-27} \\
   \lcomp{B_{cba}} = C_{cba} \label{eq:3-28}
\end{gather}
If $\Omega$ is transformed outside of its WZ gauge, non-zero components begin
to occur in \eqref{eq:3-27}. Thus, WZ gauge for the prepotential corresponds
to the gauge one would naturally define for the coefficients $B_{CBA}$. We will
assume, henceforth, that we are working in the WZ gauge.

The component fields of the three-form supermultiplet are defined as follows.
We write
\begin{equation}
   \lcomp{B_{mnp}} = C_{mnp}
\label{eq:3-29}
\end{equation}
while
\begin{equation}
\begin{gathered}
   \lcomp{Y} = Y \; , \quad  \lcomp{\bar{Y}} = \bar{Y} \\
   \lcomp{\D_\a Y} = \sqrt{2}\eta_\a \; , \quad
       \lcomp{\Db_\ad Y} = \sqrt{2}\bar{\eta}_\ad \\
\end{gathered}
\label{eq:3-30}
\end{equation}
and
\begin{equation}
   \lcomp{\D^2 Y} + \lcomp{\Db^2\bar{Y}} = -8H
\label{eq:3-31}
\end{equation}
Equation \eqref{eq:3-19} means that the orthogonal combination is not
an independent component field. It is given by the expression
\begin{multline}
   \lcomp{\D^2 Y} - \lcomp{\Db^2\bar{Y}} = 
       \frac{32i}{3}\epsilon^{mnpq}\partial_{m}C_{npq}
       + 2\sqrt{2}i \bar{\psi}_{m}\bar{\s}^{m}\eta \\
       - 2\sqrt{2}i \psi_{m}\s^{m}\bar{\eta}
       - 4 \left( \bar{M}+\bar{\psi}_{m}\bar{\s}^{mn}\bar{\psi}_{n} \right) Y 
       + 4 \left( M+\psi_{m}\s^{mn}\psi_{n} \right) \bar{Y} 
\label{eq:3-32}
\end{multline}

Under a combination of a general superdiffeomorphism, $\delta_{\xi}$, and 
an arbitrary three-form supergauge transformation, $\delta_{\Lambda}$, 
the super three-form transforms as
\begin{equation}
   \left(\delta_{\xi}+\delta_{\Lambda}\right) B = 
        \iota_{\xi}\Sigma + d\left(\Lambda+ \iota_{\xi}B\right)
\label{eq:3-33}
\end{equation}
where $\iota_{\xi}$ is the super inner derivative. The
superdiffeomorphism is parameterized by a general supervector field
$\xi^A$, while the gauge transformation parameter is a super two-form
$\L$. Clearly, such a transformation generically takes $B$ out of the WZ gauge.
However, as in the flat-space case, it is possible to find a
compensating gauge transformation $\L_\xi$, dependent on the
parameter $\xi^A$, such that the combined transformation leaves $B$ in
WZ gauge. We now restrict ourselves to the diffeomorphisms
corresponding to the supersymmetry transformations of the supergravity
multiplet \eqref{eq:3-4}. It is possible to show from the results of
\cite{BPGG} that the corresponding transformation of the super
three-form, restricted to WZ gauge, gives for the component $C_{mnp}$,
\begin{equation}
   \left(\delta_{\xi}+\delta_{\Lambda_{\xi}}\right) C_{mnp} =
       \frac{\sqrt{2}}{16} \left( \bar{\xi}\bar{\sigma}^{q}\eta
            - \xi\s^q\bar{\eta} \right) \epsilon_{qmnp}
       - \frac{3}{4} \left[ \left(\psi_{[m}\sigma_{np]}\xi\right) \bar{Y}
            + \left(\bar{\psi}_{[m}\bar{\sigma}_{np]}\bar{\xi}\right) Y \right]
\label{eq:3-36}
\end{equation}
%
%
%

Up to this point, we have been discussing the theory of an independent 
three-form superfield coupled to old minimal supergravity. 
As we now show, by the application of two
additional constraints, these two multiplets can be merged into one
irreducible supergravity multiplet containing the real three-form component
field. To do this, we impose two new constraints 
\begin{equation}
   \Si_{\d cba} = {\Si^\dd}_{cba} = 0
\label{new_constraint}
\end{equation}
which imply 
\begin{equation}
   \bar{\cal{D}}^{\dot{\alpha}}\bar{Y}=0 \; , \quad 
   {\cal{D}}_{\alpha}Y=0
\label{eq:3-38}
\end{equation}
These constraints, in addition to the original chirality constraints in
\eqref{eq:3-18}, mean that superfield $Y$ must be a constant.
We will choose superfield $Y=y$.
It follows immediately from expressions
\eqref{eq:3-30} and \eqref{eq:3-31} that component fields are
\begin{equation}
   Y = y 
\label{eq:3-39}
\end{equation}
and
\begin{equation}
   \eta_{\alpha} = H = 0
\label{eq:3-40}
\end{equation}
Thus, the additional constraints \eqref{eq:3-38} eliminate all the component
fields coming from the three-form supermultiplet except for $C_{mnp}$.
However, these constraints do something more interesting. It follows
from expression \eqref{eq:3-32} that now
\begin{equation}
   M_{2} = -\frac{3}{4y}\epsilon^{mnpq}\partial_{m}C_{npq} 
       + \frac{i}{2} \left(
           \psi_m\s^{mn}\psi_n-\bar{\psi}_m\bar{\s}^{mn}\bar{\psi}_n 
           \right)
\label{eq:3-41}
\end{equation}
That is, component field $M_{2}$ of the original supergravity multiplet is
also eliminated in terms of the three-form and gravitino component fields.
Furthermore, one should note that if we choose constant $y=-4/3$, then
the normalization of this expression matches that of equation
\eqref{eq:3-7} which was found to be the appropriate relation to insure
integrability of the supersymmetry transformation law. Henceforth, we will
take $y$ to have this value. The remaining component
fields of the original supergravity multiplet, ${e_m}^a$, 
${\psi_m}^\a$, $b_{a}$ and $M_{1}$ are 
unaffected by the new constraints. Their supersymmetry
transformations are still given by the first three equations in \eqref{eq:3-4}
and equation \eqref{eq:3-5} where, however, $M_{2}$ must be written in terms of
$C_{mnp}$ using \eqref{eq:3-41}. Transformation \eqref{eq:3-6} is no longer
relevant. It is now replaced by supersymmetry transformation \eqref{eq:3-36}.
Using \eqref{eq:3-39} and \eqref{eq:3-40}, this transformation becomes
\begin{equation}
   \left(\delta_{\xi}+\delta_{\Lambda}\right) C_{mnp} = 
        \frac{i}{3}\epsilon_{mnpq} \left(
             \xi\s^{qr}\psi_r-\bar{\xi}\bar{\s}^{qr}\bar{\psi}_{r} 
             \right)
 \label{eq:3-42}
\end{equation}
Note that this is identical to the integrated supersymmetry transformation for
$C_{mnp}$ given in expression \eqref{eq:3-8} if we identify
\begin{equation}
   \delta = \delta_{\xi}+\delta_{\Lambda_\xi}
\label{eq:3-43}
\end{equation}

The new constraints are superfield equations. Thus the supersymmetry
transformations for the, now eliminated, component fields $Y$,
$\eta_{\alpha}$ and $H$ are automatically consistent, all being
identically zero. We conclude that the additional constraints
\eqref{eq:3-38} lead to a supergravity multiplet consisting of
the component fields ${e_m}^a$, ${\psi_m}^\a$, $b_{a}$, $M_{1}$ and
$C_{mnp}$. These fields transform irreducibly under the supersymmetry
transformations given in equations \eqref{eq:3-9}. Thus, we have found a
complete, off-shell superfield representation of the three-form 
supergravity multiplet.

For completeness let us summarize the full set of superfield
constraints we used to derive this multiplet. We started in curved
superspace with a supervielbein ${E_M}^A$ and a Lorentz-algebra-valued
connection ${\phi_A}^B$, together with a super three-form potential
$B$. For the vielbein we adopted the usual torsion constraints of old
minimal supergravity, as given, for instance, in \cite{WessBagger}, 
\begin{equation}
\begin{gathered}
   {T_{\underline{\a\b}}}^{\underline{\g}} = 
   {T_{\a\b}}^c = {T_{\ad\bd}}^c = 
   {T_{\underline{\a}b}}^c = {T_{a\underline{\b}}}^c =
   {T_{ab}}^c = 0 \\
   {T_{\a\bd}}^c = {T_{\bd\a}}^c = 2i{\s^c}_{\a\bd}
\end{gathered}
\label{Tconstraints}
\end{equation}
Now we turn to the super form-form field strength $\Si=dB$. From
equations \eqref{eq:3-13}--\eqref{eq:3-17} and \eqref{new_constraint},
with the normalization $y=-4/3$, we see that all the components of
$\Si$ are set to zero except for 
\begin{equation}
   \Sigma_{\a\b ba} = 
      - \frac{2}{3}\left(\s_{ba}\e\right)_{\a\b}  \quad
   {\Sigma^{\ad\bd}}_{ba} = 
      - \frac{2}{3} \left( \bar{\s}_{ba}\e\right)^{\ad\bd} 
\label{sigma_constraints}
\end{equation}
and the component $\Si_{abcd}$, which is unconstrained. Solving the
$d\Si=0$ Bianchi identity does, however, determine $\Si_{abcd}$ in terms of
the curvature superfield $R$. From \eqref{eq:3-19} we see
\begin{equation}
   \Si_{abcd} = \frac{i}{2} \left( R^\dag - R \right) \e_{abcd}
\end{equation}
which links the vielbein and three-form field into a single
irreducible supergravity multiplet. 

We end this section by computing the commutator of the transformations
$\delta$ acting on the component field $C_{npq}$. We find that
\begin{equation}
   \left(\delta\delta'-\delta'\delta\right) C_{mnp} = 
      2\delta_\eta C_{mnp} + \delta_{\xi''}C_{mnp} + \d_{\L''} C_{mnp}
\label{eq:3-44}
\end{equation}
where we define
\begin{align}
   \delta_\eta C_{mnp} &= 
      \eta^{q}\p_{q}C_{mnp} + \left(\p_{m}\eta^{q}\right)C_{qnp}
      + \left(\p_{n}\eta^{q}\right)C_{mqp} 
      + \left(\p_{p}\eta^{q}\right)C_{mnq} \\ \label{eq:3-45}
   \d_{\xi''}C_{mnp} &= \frac{i}{3}\epsilon_{mnpq}
      \left(\xi''\sigma^{qr}\psi_{r} 
         - \bar{\xi''}\bar{\sigma}^{qr}\bar{\psi}_{r}\right) \\
   \d_{\L''}C_{mnp} &= 3 \p_{[m} \L''_{np]}
\end{align}
The fermionic parameter $\xi''$ and the two-form $\L''_{mn}$ are given by
\begin{align}
   {\xi''}^\a &= \eta^m \psi_m^{\ \a} \\
   \Lambda''_{mn} &= -3C_{mnp}\eta^{p} - \frac{1}{3}\eta^{+}_{mn}
\label{eq:3-46}
\end{align}
while the parameters $\eta^{r}$ and $\eta^{+}_{pq}$ are given in
\eqref{eq:2-37}. What is the interpretation of the three terms on the
right hand side of equation \eqref{eq:3-44}? Substituting
$\delta=\delta_{\xi}+\delta_{\Lambda_\xi}$ on the left side of
\eqref{eq:3-44},  and using the fact that
$\d_{\L_\xi}\d_{\xi'}C_{mnp}=\d_{\L_\xi}\d_{\L'_\xi}C_{mnp}=0$, we find 
\begin{equation} 
   \left(\delta_\xi\delta_{\xi'}-\delta_{\xi'}\delta_\xi\right) C_{mnp} =
      2\delta_\eta C_{mnp} + \delta_{\xi''}C_{mnp}
\label{eq:3-47}
\end{equation}
and
\begin{equation}
   \left( \d_{\xi}\d_{\L'_\xi} - \d_{\xi'}\d_{\L_\xi} \right)C_{mnp} =
      \d_{\L''}C_{mnp}
\label{eq:3-48}
\end{equation}
From equation \eqref{eq:3-47} we see that $\d_\eta C_{mnp}$ and
$\d_{\xi''}C_{mnp}$ come from the superdiffeomorphism algebra. They
may be understood in the following way. The term $\d_\eta C_{mnp}$ is
just a usual bosonic diffeomorphism; $\d_{\xi''}C_{mnp}$ is, in fact,
a local supersymmetry transformation with parameter $\xi''$. These are
the restricted class of transformations which leave the supervielbein
in the gravitational WZ gauge. Equation \eqref{eq:3-47} is just
reproducing the well-known algebra of these restricted
diffeomorphisms. The fact that it is not possible to define a local
supersymmetry transformation which closes just into a bosonic
diffeomorphism is a result of the requirement that one remains in the
gravitational WZ gauge. 

Turning to equation \eqref{eq:3-48} we see that the $\d_{\L''}C_{mnp}$
term arises because we have also chosen a WZ gauge for the three-form
$B$. If WZ gauge restoration was unnecessary, we would have
$\d=\d_\xi$ and the first two terms in \eqref{eq:3-44} would be 
the only terms to appear. However, WZ gauge restoration is
necessary and \eqref{eq:3-48} tells us that the effect of this gauge
restoration is the third term on the right hand side of \eqref{eq:3-44}. 
Note that the term proportional to $\eta^{+}_{pq}$ in $\Lambda$ is field
independent. Therefore, it would appear to be a very different situation 
than in the previous section. However, one should note that this term
is, in fact, identical to the second term on the right side of
equation \eqref{eq:2-39} with $Y_{1}=y=-4/3$ and $Y_{2}$ set to zero,
as is required by \eqref{eq:3-39}. 

As in the flat-space case, one might have been tempted to
interpret this term as indicating the presence of central extension to
the local supersymmetry algebra associated with supermultiplet
$\Omega$. The suggestion is that, in general, two superdiffeomorphisms
can close in to a gauge transformation $\d_{\L''}$. It is clear from
the above analysis that this interpretation is unwarranted. The
general symmetry of the multiplet is local superdiffeomorphisms, which
form a closed subalgebra, together with independent gauge
transformations. The closure of two supersymmetry transformations into
a gauge transformation is an artifact of the three-form WZ gauge.

We have carried out this analysis by re-expressing the imaginary part
of the auxiliary field $M$ in terms of the three-form $C_{npq}$. In
terms of the component field calculation we could just as easily
dualized the real part of $M$ or some linear combination of  $M_1$ and
$M_2$. Indeed, $M$ could have been completely replaced by two
independent three-forms. From the superfield point of view, only one 
three-form is introduced. The freedom in choosing a particular linear 
combination of auxiliary fields to dualize is reflected in the arbitrariness 
in choosing the phase in the constant $y$. However, as we will see in section 
five, demanding the Green-Schwarz membrane action is $\k$-symmetric, picks 
out specifically the $M_2$ dualization.


\section{The elementary membrane solution}


We claim that the three-form version of $N=1$ supergravity derived in
the previous section is the version relevant to the description of
fundamental membranes in four dimensions. In the next section, we show
that $\kappa$-symmetry of the Green-Schwarz membrane action leads to
exactly the superfield constraints used in section three. This section
demonstrates how the fundamental membrane solution appears in the 
three-form supergravity theory. 

Since we are now interested in solutions rather than supergravity
multiplets, we must begin by giving the action for the three-form
supergravity. As discussed above, the old minimal supermultiplet of 
$D=4$, $N=1$ supergravity 
consists of a graviton $g_{mn}$, a gravitino $\psi_{m}^{\ \alpha}$, a real
vector field $b_a$ and a complex scalar field $M$. The bosonic part of the
pure supergravity Lagrangian is given by
\begin{equation}
   \kappa^{2}e^{-1}{\cal{L}} = 
       \frac{1}{2}{\cal{R}} + \frac{1}{3}b^2
       - \frac{1}{3}M_{1}^{2}-\frac{1}{3}M_{2}^{2}
\label{eq:4-1}
\end{equation}
where $M=M_{1}+iM_{2}$, The three-form supergravity multiplet was
derived by dualizing $M_2$ into a four-form field strength
$F_{mnpq}$. If we continue to ignore fermion terms, the transformation
is 
\begin{equation}
   M_2 = \frac{1}{4}\e^{mnpq}F_{mnpq} = \e^{mnpq}\p_m C_{npq}
\label{eq:4-4}
\end{equation}
where we write $\lcomp{\Si_{mppq}}=F_{mnpq}$. Substituting into the
supergravity Lagrangian \eqref{eq:4-1} gives
\begin{equation}
   \kappa^{2}e^{-1}{\cal{L}} = \frac{1}{2}{\cal{R}}
       + \frac{1}{3}b^2 - \frac{1}{3}M_{1}^{2} + 2F^{2}
\label{eq:4-3}
\end{equation}

This lagrangian is not quite equivalent to the original
one. Varying the auxiliary fields in \eqref{eq:4-1} leads to the field
equations $M_1=M_2=b_a=0$. In \eqref{eq:4-3}, $b_a$ and $M_1$
similarly get set to zero, but the $C_{mnp}$ field equation gives
\begin{equation}
   \nabla^m F_{mnpq} = 0 \quad \Leftrightarrow \quad F_{mnpq} = \l\e_{mnpq}
\label{Ceqn}
\end{equation}
where $\l$ is a constant of integration. Thus, from \eqref{eq:4-4}, we
see that now $M_2=\frac{1}{4}\e^{mnpq}F_{mnpq}=-6\l \neq 0$. In fact,
the constant of integration $\l$ acts as a cosmological constant. On
shell, where $b_a=M_1=0$, the  metric equation of motion reads
\begin{equation}
   R_{mn} - \frac{1}{2}g_{mn}R = 
       - 4{F_m}^{pqr}F_{npqr} - \frac{1}{2}g_{mn}F^2 = 12 \l^2 g_{mn}
\label{eq:4-32}
\end{equation}
where we have substituted the solution of the $C_{mnp}$ equation
\eqref{Ceqn}. We see that \eqref{eq:4-32} is simply Einstein's
equation for empty space with a negative cosmological constant
$\L=-12\l^2$. Note that the value of $\L$ is determined dynamically by
solving the $C_{mnp}$ equation of motion. 

The equivalence of a three-form field coupled to gravity and a
dynamically determined cosmological constant was noted some time ago
in \cite{ANT,DN,Cdual}. In the present context, it provides a way of
deriving the fermion terms that had to be added to $M_2$ in the
definition \eqref{eq:3-7} to make the supersymmetry variation of
$F_{mnpq}$ integrable. The full action for old-minimal supergravity
with a cosmological constant is 
\begin{multline}
   S = \frac{1}{2\k^2} \int{d^4x}\; \sqrt{-g} \left\{ R 
        + \e^{mnpq}\left( \psib_m\sb_n\tilde{\D}_p\psi_q
            + \psi_m\s_n\tilde{\D}_p\psib_q \right)
        + \frac{2}{3}b^2 - \frac{2}{3}\left( {M_1}^2 + {M_2}^2 \right)
        \right. \\ \left.
        + \m_1 \left[ M_1 + \frac{1}{2}\left( \psi_a\s^{ab}\psi_b 
            + \psib_a\sb^{ab}\psib_b \right) \right]
        + \m_2 \left[ M_2 - \frac{i}{2}\left( \psi_a\s^{ab}\psi_b 
            - \psib_a\sb^{ab}\psib_b \right) \right] \right\}
\label{cosmo_action}
\end{multline}
Eliminating $M_1$ and $M_2$, we find the cosmological constant is
$\L=-\frac{3}{16}\left({\m_1}^2+{\m_2}^2\right)$. We would like to
replace $M_2$ by a three-form field such that the cosmological constant
is fixed by the three-form equation of motion. Since $M_1$ is not
dualized, we first take $\m_1=0$. Defining 
\begin{equation}
   \frac{1}{4}\e^{mnpq}F_{mnpq} = \e^{mnpq}\p_m C_{npq} = M_{2}
       - \frac{i}{2}\left( \psi_a\s^{ab}\psi_b 
          - \bar{\psi}_a\bar{\s}^{ab}\bar{\psi}_b \right)
\label{eq:4-2}
\end{equation}
means that the last term in \eqref{cosmo_action} is a total
divergence. The parameter $\m_2$ effectively drops from the action and
the cosmological constant becomes fixed dynamically by solving the
$C_{mnp}$ equation of motion. The definition \eqref{eq:4-2} is exactly
the combination that appeared in the supersymmetric dualization of
$M_2$ in the previous section \eqref{eq:3-7}. One notes that, since the 
action with $\m_1=\m_2=0$ is supersymmetric, the last term in \eqref{eq:4-2} 
must be invariant by itself. Thus the variation of $\e^{mnpq}F_{mnpq}$ must
be a total divergence and so can be integrated to give a variation of
$C_{mnp}$. This was just the property required in the previous section
to form the three-form multiplet. 

Since the total-derivative $\m_2$ term can effectively be dropped, if
we rewrite $M_2$ in terms of $C_{mnp}$ and put the auxiliary $b_a$ and
$M_1$ fields on shell, the bosonic lagrangian for three-form
supergravity becomes
\begin{equation}
   S = \frac{1}{2\k^2}\int{d^{4}x}\sqrt{-g} \left( {\cal R} + F^2 \right)
\label{eq:4-5}
\end{equation}
As we have argued, this is classically equivalent to pure gravity with
a negative cosmological constant $\L=-12\l^2$ where
$F_{mnpq}=\l\e_{mnpq}$ is the solution of the $C_{mnp}$ equation of
motion. 

It is interesting to note that the action has the same general structure as
higher-dimensional supergravity actions, such as type IIA supergravity
in $D=10$ or $D=11$ supergravity, with Einstein gravity coupled to
Neveu-Schwarz and Ramond forms. There is, of course, no dilaton field
in our $D=4$ action. More importantly, the sign in front of the $F^{2}$ term 
in \eqref{eq:4-5} positive, whereas the usual sign in front of the form-field 
kinetic terms in higher-dimensional supergravity actions is negative. 

We would like to find a fundamental membrane solution which is a source
of three-form charge. The three-dimensional membrane is a domain
wall, dividing the spacetime into two regions. Domain wall solutions
for general $(D-2)$-branes in $D$-dimensional spacetime have been
discussed in \cite{LPSS,LPT,CLPST}. Here we reproduce the solution in
terms of a three-form and map to an explicit membrane source. 

Let the spacetime coordinates be $t$, $x$, $y$ and $z$. If a flat 
membrane lies in the $t$--$x$--$y$ plane, its Lorentz and
translational symmetries imply that the spacetime metric has the form
\begin{equation}
   ds^{2} = e^{2A(z)} \left( - dt^2 + dx^2 + dy^2 \right) + dz^2
\label{eq:4-11}
\end{equation}
Since the three-form action \eqref{eq:4-5} is equivalent to the action
for pure gravity with a negative cosmological constant, the
symmetries of \eqref{eq:4-11} are sufficient to determine the
solution. The spacetime is simply anti-de Sitter space. The membrane
solution then corresponds to patching of two pieces of anti-de Sitter space,
as was first described in \cite{LPT}. 

The details of the solution are as follows. The source is described by
the bosonic action for a fundamental membrane coupled to $C_{mnp}$
\begin{equation}
   S_{3} = T_{3} \int{d^{3}\xi} \sqrt{-\gamma}\left( 
       - \frac{1}{2}\gamma^{ij}\partial_{i}X^{m}\partial_{j}X^{n}g_{mn}
       + \frac{1}{2}
       - \frac{{\cal{C}}}{6}\e^{ijk}\p_{i}X^{m}\p_{j}X^{n}\p_{k}X^{p}C_{mnp}
       \right)
\label{eq:4-6}
\end{equation}
where $T_{3}$ is the membrane tension, $\gamma_{ij}$ is the intrinsic
membrane metric, $\e_{ijk}$ is the volume three-form on the membrane
and the $X^{m}$ fields describe the embedding of the membrane into the
four-dimensional target space. The dimensionless  constant ${\cal{C}}$
is not known a priori but, for a consistent solution, will be fixed by
the normalization of the $F^2$ term in \eqref{eq:4-5} in what follows.
The full action is then the sum of
\eqref{eq:4-5} and \eqref{eq:4-6}. The associated field equations are
the following. The metric field equation is 
\begin{equation}
   {\cal{R}}^{mn} - \frac{1}{2}g^{mn}{\cal{R}} 
      + 4{F^m}_{pqr}F^{npqr} - \frac{1}{2}g^{mn}F^2
   = \kappa^{2}T^{mn}
\label{eq:4-7}
\end{equation}
where $T^{mn}$ is the membrane stress-energy tensor given by
\begin{equation}
   T^{mn} = -T_{3} \int{d^{3}\xi} \sqrt{-\gamma}\gamma^{ij}
       \p_i X^m \p_j X^n \frac{\d^{4}\left(x^{p}-X^{p}\right)}{\sqrt{-g}}
\label{eq:4-8}
\end{equation}
The three-form field equation is found to be
\begin{equation}
   \p_{m}(\sqrt{-g}F^{mnpq}) = 
       -2\kappa^{2}T_{3}{\cal{C}}\int{d^{3}\xi} \sqrt{-\gamma}
           \e^{ijk}\p_i X^n \p_j X^p \p_k X^q \d^{4}\left(x^{p}-X^{p}\right)
\label{eq:4-9}
\end{equation}
and, finally, the $X^{m}$ equation of motion is
\begin{multline}
   \qquad \p_{i}\left(\sqrt{-\g}\g^{ij}\p_{j}X^{n}g_{qn}\right)
        - \frac{1}{2}\sqrt{-\g}\g^{ij}\p_{i}X^{m}\p_{j}X^{n}\p_{q}g_{mn}
        \\
        - \frac{{\cal{C}}}{6}\sqrt{-\g}
            \e^{ijk}\p_{i}X^{m}\p_{j}X^{n}\p_{k}X^{p}F_{qmnp}
   = 0
\label{eq:4-10}
\end{multline}
while the membrane-metric equation gives
$\gamma_{ij}=\partial_{i}X^{m}\partial_{j}X^{n}g_{mn}$. 

As discussed above, the metric Ansatz for a membrane lying in the
plane $z=0$ is given by \eqref{eq:4-11}. Adopting a gauge where the
three-form shares the symmetries of the membrane, means the Ansatz for
$C_{mnp}$ is that all components vanish except for 
\begin{equation}
   C_{txy} = e^{C(z)}
\label{eq:4-12}
\end{equation}
For the flat membrane itself, we choose static gauge where $T=\s^0$,
$X=\s^1$ and $Y=\s^2$. The Einstein equations imply that
\begin{gather}
   2\p_z^2 A + 3\left(\p_z A\right)^2 - 12e^{-6A+2C} \left(\p_z C\right)^2
      = -\k^2 T_3 \d (z) \label{eq:4-17} \\
   \frac{1}{3}e^{-6A} \p_z \left( e^{3A} - 6e^C \right)
      \p_z \left( e^{3A} + 6e^C \right) = 0 \label{eq:4-18}
\end{gather}
whereas the three-form equation and the $X^{m}$ equation give
\begin{equation}
   \p_z \left( e^{-3A} \p_z e^{C} \right) = 
      - \frac{\kappa^{2}}{24}T_{3}{\cal{C}} \d (z)
\label{eq:4-19}
\end{equation}
and
\begin{equation}
   \p_z \left(e^{3A}-{\cal{C}}e^{C}\right) = 0
\label{eq:4-20}
\end{equation}
respectively. From equation \eqref{eq:4-18} and \eqref{eq:4-20} we see
that
\begin{equation}
   e^{C} = \pm\frac{1}{6}e^{3A} + \text{const} \; , \qquad
   {\cal C} = \pm 6
\label{eq:4-25}
\end{equation}
The constant can be set to zero by a gauge transformation. We see that
consistency of the membrane and spacetime equations of motion fixes
the magnitude of $\cal C$. The particular value is fixed by our choice of
normalization of $F^2$ in \eqref{eq:4-5}. The sign of
$\cal C$ will be related in the next section to a choice of
$\k$-symmetry helicity on the membrane. To match the conventions
there, we will henceforth take ${\cal C}=-6$. 

The above equations of motion now all reduce to 
\begin{equation}
   \p_z^2 A =  -\frac{\kappa^{2}}{2}T_{3}\d (z)
\label{eq:4-23}
\end{equation}
The solution symmetrical in $z$ is
\begin{equation}
   A = -\a \left| z \right|
\label{eq:4-29}
\end{equation}
where $\a=\frac{1}{4}\k^2 T_3$. Given the factor of $\sqrt{-g}$ in
$\e_{mnpq}$, we can write the final solution as
\begin{equation}
\begin{gathered}
   ds^2 = e^{-\a |z|} \left( - dt^2 + dx^2 + dy^2 \right) + dz^2 \\
   F_{txyz} = \pm \frac{1}{2} \e_{txyz}
\end{gathered}
\end{equation}
where one takes the $-$ or $+$ sign in the expression for $F_{txyz}$
depending on whether one is the right ($y>0$) or left ($y<0$) of the
membrane. As expected, the four-form field is proportional to
$\e_{mnpq}$. From \eqref{eq:4-32} we expect the metric to correspond to
anti-de Sitter space with a cosmological constant
$\L=-3\a^2=-3\k^4T_3^2/16$. This is the case; the metric is written in
unconventional horospherical coordinates \cite{adSmetric}.

Thus there is a fundamental membrane solution of three-form
supergravity which matches consistently onto a membrane source. The
solution is the patching of two pieces of anti-de Sitter space across a
singular plane. The four-form field strength is constant either side
of the membrane, but with opposite sign. This corresponds to the field
on each side `pointing away' from the source. The four-form cannot
fall off away from the membrane because the transverse space is only
one-dimensional. Solutions for multiple parallel branes are, of
course, easily constructed, as discussed in \cite{LPT}. 

Finally, again following \cite{LPT}, let us briefly discuss the
supersymmetry of the membrane solution. It is well known that anti-de
Sitter space is supersymmetric, so that, away from the membrane
source, we expect there to be four distinct Killing spinors. The
supersymmetric variation of the gravitino is given by 
\begin{equation}
   \d {\psi_m}^\a = - 2\p_m\xi^\a - \o_{mnp} \left(\e\s^{np}\xi\right)^\a
       - \a \left(\e\s_m\bar{\xi}\right)^\a
\label{eq:4-40}
\end{equation}
The spin connection $\omega_{mnl}$ can be evaluated from the vielbein
associated with metric \eqref{eq:4-11} and vanishes except for 
\begin{equation}
   \o_{ttz} = -\o_{xxz} = -\o_{yyz} = e^{2A} \p_z A
\label{eq:4-41}
\end{equation}
The Killing spinors are determined by setting the variation of the gravitino 
to zero for each value of index $m$. First consider the $m=z$
case. The spin-connection term vanishes in this equation. The solution is 
a spinor of the form
\begin{equation}
   \xi^\a = e^{-A/2} \eta^{+\a} + e^{A/2} \eta^{-\a}
\label{eq:4-42}
\end{equation}
where we define 
\begin{equation}
   \eta^\pm = \eta \pm \s_3 \etab
\end{equation}
and the general spinor $\eta$ is an arbitrary function of $t$, $x$ and
$y$. Setting the remaining components of the gravitino variation to
zero fixes the form of $\eta$. After some  Dirac matrix
manipulation, we find the general solution 
\begin{equation}
   \xi = \left(e^{-A/2}\eta^+_0 
          \mp \frac{\a}{2} e^{A/2} x^i\s_i \etab^+_0\right)
      + e^{A/2} \eta^-_0 
\label{eq:4-44}
\end{equation}
where $i=\{t,x,y\}$ and the $-$ or $+$ sign depends on whether one is
to the right or the left of the membrane. As before
\begin{equation}
   \eta^\pm_0 = \eta_0 \pm \s_3 \etab_0 
\label{projspinor}
\end{equation}
where now $\eta_0$ is an arbitrary constant spinor. Since $\eta_0$ has four
real components, there are four independent Killing spinors 
described by this expression. We see that
the solution naturally decomposes with respect to the two chiralities
of $\eta_0$ given by \eqref{projspinor}. While both chiralities are
Killing spinors of the anti-de Sitter space, it is not clear that both
are compatible with the singular membrane surface at $z=0$. 

In fact only the two $\eta^-_0$ chirality spinors are
compatible. There are several ways to see this. The simplest, and the
one we employ here, is to note that the corresponding Killing spinors,
given by 
\begin{equation}
   \xi = e^{A/2} \eta^-_0
\label{eq:4-56}
\end{equation}
are continuous at the membrane surface. We can can conclude, therefore,
that these Killing spinors are globally defined. For the two spinors of
the opposite chirality, the corresponding Killing spinors are
given by  
\begin{equation}
   \eta = e^{-A/2}\eta^+_0 \mp \frac{\a}{2} e^{A/2} x^i\s_i \etab^+_0
\label{eq:4-58}
\end{equation}
Note that the sign of the second term changes as one passes across the
membrane surface. Thus the Killing spinors are discontinuous and are
not compatible with the membrane source. We conclude
that the membrane preserves precisely half of the supersymmetries
and, hence, is a BPS solution. Of the four Killing spinors
given above, only the chiral spinors \eqref{eq:4-56} are global Killing
spinors of the complete membrane solution.


\section{Membrane $\kappa$-symmetry and the supergravity constraints}


In the previous section, we demonstrated that the dual form of 
supergravity, in 
terms of the three-form potential, allowed a fundamental BPS membrane 
solution. In this section, we strengthen this link between membranes and 
the dual form of supergravity by considering the fundamental supersymmetric 
Green-Schwarz membrane. We show that the theory is only consistent in
curved superspace if there exists a super three-form field satisfying
exactly the constraints of the three-form supergravity given in
section two. The consistency condition arises by insisting that the
membrane action has a $\kappa$-symmetry, implying that half the
fermionic degrees of freedom decouple on shell, and that the membrane
consequently has the same number of fermionic as bosonic
excitations. The argument was first given by Bergshoeff, Sezgin and
Townsend in \cite{BST}. These authors presented a general formalism
for $n$-extended objects in $d$-dimensional supergravity
backgrounds. They argued that, for membranes, the only relevant
supergravities are $N=1$, $D=11$ and $N=2$, $D=7$. Since the
superspace formalism of $N=2$, $D=7$ is unknown, they concentrated on
the case of a supermembrane in eleven dimensions. However, the $N=1$,
$D=4$ three-form supergravity presented in this paper gives yet
another context in which a supermembrane can couple to
supergravity. In this section, we apply the formalism of \cite{BST} to
this four-dimensional case. 

The supersymmetric Green-Schwarz action for a membrane moving in curved 
superspace is given by 
\begin{equation}
   S = T_3 \int d^3\xi\sqrt{-\g} \left(
       - \frac{1}{2}\g^{ij}{E_i}^a{E_j}^b\eta_{ab} + \frac{1}{2}
       + \e^{ijk}{E_i}^A {E_j}^B {E_k}^C B_{CBA}
       \right)
\label{GSaction}
\end{equation}
The target space is now the full four-dimensional superspace so that the 
membrane coordinates are $Z^M=(X^m(\xi^i),\q^\a(\xi^i),\qb_\ad(\xi^i))$. 
The world volume metric on the membrane is $\g_{ij}$ while the functions 
${E_i}^A$ are given by 
\begin{equation}
   {E_i}^A = \p_i Z^M {E_M}^A
\end{equation}
where ${E_M}^A$ are the usual inverse supervielbeins. The action also 
includes a Wess-Zumino term describing the coupling of the membrane to 
some, as yet unconstrained, super three-form field $B_{CBA}$. The bosonic 
part of this action was given in equation \eqref{eq:4-6} in the previous 
section, with the ${\cal C}=-6$, as we derived for the consistent
membrane solution. 

As written, the action \eqref{GSaction} is spacetime supersymmetric by 
construction. However, to describe equal numbers of fermionic and bosonic 
degrees of freedom it must also have a $\kappa$-symmetry. The
existence of this symmetry requires a non-zero Wess-Zumino term in
\eqref{GSaction}. Thus the membrane theory only exists in a non-zero
three-form background. Following \cite{BST}, $\k$-symmetry implies
that the action is invariant under a variation of the form 
\begin{equation}
\begin{aligned}
   \d E^a &= 0 \\
   \d E^\a &= \k^\a + \kb_\ad {\bar\G}^{\ad\a} \\
   \d E_\ad &= \kb_\ad + \k^\a \G_{\a\ad} \\
   \d \g_{ij} &= 2 \left( X_{ij} - \g_{ij} X_k^{\ k} \right)
\end{aligned}
\label{k-variation}
\end{equation}
where $\d E^A=\d Z^M {E_M}^A$. Here $X_{ij}$ is a function of the 
${E_i}^A$ and is linear in $\kappa$, while 
the matrices $\G$ and $\bar\G$ are given by
\begin{equation}
\begin{aligned}
   {\bar\G}^{\ad\a} &= \frac{1}{6}\e^{ijk}E_i^{\ a}E_j^{\ b}E_k^{\ c}
      \e_{abcd} \bar\s^{d\ad\a} \\
   \G_{\a\ad} &= -\frac{1}{6}\e^{ijk}E_i^{\ a}E_j^{\ b}E_k^{\ c}
      \e_{abcd} \s^d_{\a\ad}
\end{aligned}
\end{equation}
In Dirac four-spinor notation we have
\begin{equation}
   \G = \left( \begin{array}{cc} 0 & {\bar\G}^{\a\bd} \\ 
           \G_{\ad\b} & 0 \end{array} \right)
\end{equation}
Using the the $\g_{ij}$ equation of motion $\g_{ij} = {E_i}^{a}{E_j}^{b}
\eta_{ab}$ it is easy to show that $1+\G$ is a projector on Dirac
spinors since $\G^2=1$. One notes that, for a flat membrane, $1+\G$ is
the same projector that appears in the Killing spinors of membrane
solution given in the previous section. In static gauge a membrane in
the plane $z=0$ has $T=\s^0$, $X=\s^1$ and $Y=\s^2$, so that
$\G_{\a\ad}={\s^3}_{\a\ad}$. This gives $\d
E_\a=\left(\k-\s^3\kb\right)^\a$ which is the same projection used for the
preserved Killing spinors \eqref{eq:4-56}. This is to be expected: it
reflects the usual realization of partially broken supersymmetry in
the flat membrane world-volume theory \cite{Polchin}. It is also the
choice of chiral projector $1\pm\G$ in the $\k$-symmetry which fixes
the sign of the $B_{CBA}$ coupling ${\cal C}=\mp 6$ discussed in the
previous section. 

For general supervielbein ${E_M}^A$ and
three-form potential $B_{CBA}$, variations of the form
\eqref{k-variation} do not leave the action \eqref{GSaction}
invariant. Requiring invariance imposes conditions on the derivatives
of the supervielbein and three-form potential. In particular,
following \cite{BST}, we find that certain components of the torsion,
constructed from ${E_M}^A$, are constrained to be 
\begin{gather}
   {T_{\a\b}}^c = {T_{\ad\bd}}^c = 0 \; , \qquad
   {T_{\a\ad}}^a = {T_{\ad\a}}^a = 2i{\cal A} {\s^a}_{\a\ad} 
       \label{torsion} \\
   \eta_{c(a}{T_{b)\a}}^c = \eta_{ab} \L_\a \; , \quad 
   \eta_{c(a}{T_{b)\ad}}^c = \eta_{ab} \bar\L_\ad \label{torsionL}
\end{gather}
where ${\cal A}$ is an arbitrary real number, while $\L_\a$ is any spinor 
superfield. Similarly, the components of the four-form field strength 
constructed from $B_{CBA}$ are also constrained. These constraints are 
\begin{gather}
   \Si_{\underline{\d\g\b}A} = {\Si^\dd}_{\g ba} = 0  \label{Szeros} \\
   \Si_{\d\g ba} = -\frac{2}{3}{\cal A}\left(\s_{ba}\e\right)_{\d\g}
       \; , \quad
   {\Si^{\dd\gd}}_{ba} = 
       -\frac{2}{3}{\cal A}\left(\bar\s_{ba}\e\right)^{\dd\gd} \label{SY} \\
   \Si_{\d cba} = \frac{i}{2}{\s^d}_{\d\dd}{\bar\L}^\dd\e_{dcba}  \; , \quad
   {\Si^\dd}_{cba} = -\frac{i}{2}\bar\s^{d\dd\d}\L_\d\e_{dcba}  \label{SDY}
\end{gather}

Let us now compare these constraints with those used in section three
to derive the three-form version of supergravity, as given in
\eqref{Tconstraints} and \eqref{sigma_constraints}. The conditions are
the same if ${\cal A}=1$ and the superfield $\L_\a$ is zero, except
that the torsion is further constrained in the supergravity. In
particular, we are missing the constraints
\begin{equation}
   {T_{\underline{\a\b}}}^{\underline{\g}} = 
   \eta_{c[a}{T_{b]\underline{\a}}}^c 
   {T_{ab}}^c = 0
\label{extra_cs} 
\end{equation}
The gravitational multiplet is described in terms of the supervielbein
$E^A$ and a superconnection ${\phi_A}^B$. However there is in general
some ambiguity in how these are defined. We note first that ${\cal A}$
can be set to unity by a simple normalization condition on the
vielbein. Furthermore, several of the missing constraints in
\eqref{extra_cs} are conventional, for instance ${T_{ab}}^c=0$. That
is to say, they can be imposed simply by redefining
${\phi_A}^B$. Finally, one notes that the Green-Schwarz action
\eqref{GSaction} does not fix the definition of the bosonic and
fermionic parts of $E^A$. In general we can redefine
\begin{equation}
\begin{aligned}
   {E'}^a &= E^b {L_b}^a \\
   {E'}^\a &= E^b {L_b}^\a + E^\b {L_\b}^\a + E_\bd L^{\bd\ad} \\
   {E'}_\ad &= E^b L_{b\ad} + E^\b L_{\b\ad} + E_\bd {L^\bd}_\ad
\end{aligned}
\end{equation}
and leave the form of the action invariant. Here ${L_b}^a$ is the
usual Lorentz transformation, but the other matrices correspond to a
general linear transformation. As first described in the context of the
string in ten dimensions \cite{E_rot}, these transformations are
in fact, sufficient to set $\L_\a$ and the remaining missing torsion
constraints to zero. Thus, up to this redefinition, the constraints of
$\k$-symmetry for the membrane action are exactly equivalent to those
used to define $N=1$ three-form supergravity. One notes that, with 
${\cal A}=1$, it is precisely the $M_2$ auxiliary field which is dualized, 
and no other linear combination of $M_1$ and $M_2$. 

Rather than give the details of the redefinitions, let us demonstrate
that $\L_\a$ can consistently be taken to zero. Assume all the usual
torsion constraints are satisfied, including $\L_\a=0$ and ${\cal
A}=1$. Comparing \eqref{SY} with \eqref{eq:3-14} and \eqref{eq:3-15}
we see that the superfield $Y$ is equal to $-4/3$. Furthermore,
solving the Bianchi identities implies that $\Si_{\a
cba}\sim\Db^\ad{\bar Y}$ and  ${\Si^\ad}_{cba}\sim\D_\a Y$ (from
\eqref{eq:3-16} and \eqref{eq:3-17}). Since $Y$ is constant these components
are zero, and so, by \eqref{SDY}, $\L_\a=0$, consistent with the
original assumption in the torsion constraints.  

We see that imposing $\kappa$-symmetry on the Green-Schwarz membrane 
action implies a set of torsion and four-form constraints that are
exactly those used to define the three-form version of
supergravity. We are forced to have a non-zero $B$. The
membrane does not couple consistently to old-minimal supergravity,
but instead couples only to the three-form version of minimal
supergravity derived in this paper.


\section{The Extended Membrane Superalgebra}


It is well know that $p$-brane states in flat space are
representations of an extended superalgebra. The extension is by a topological
tensor central charge corresponding to the form-field charge of the
$p$-brane state \cite{AGITb}. In the case of a membrane in flat
four-dimensional space, we would expect the algebra to have the form 
\begin{equation}
\begin{aligned}
   \big\{ Q_\a, \Qb_\ad \big\} &= 2 {\s^m}_{\a\ad} P_m \\
   \big\{ Q_\a, Q_\b \big\} &= 2i \left(\s^{mn}\right)_{\a\b} Z_{mn} \\
   \big\{ \Qb_\ad, \Qb_\bd \big\} &= 
      2i \left(\sb^{mn}\right)_{\ad\bd} Z_{mn} 
\end{aligned}
\label{Zalgebra}
\end{equation}
where $Z_{mn}$ is the conserved charge for the three-form potential 
$C_{mnp}$. 

Normally the extended algebra is derived in one of two ways. Either
one looks at the supersymmetry algebra in the world-volume theory of a
fundamental Green-Schwarz $p$-brane \cite{AGITb}, or one looks
directly at the algebra of charges for a $p$-brane solitonic supergravity
solution, as in, for instance \cite{AGITa,DGHR}. In each case the extension
appears as a topological charge. 

In this section we will show that the algebra can
also be derived simply as the algebra of symmetries which leave the
flat-space supergravity background, including the three-form $B$,
invariant. It will be crucial in this context that the constraints on
the field strength $\Si$ imply that even in flat space there is a
non-trivial potential $B$.  

Let us start by reconsidering the algebra of general local invariances of
the supergravity derived in section three. The theory has three quite
distinct symmetries: general superdiffeomorphic invariance, local
Lorentz symmetry and symmetry under local gauge transformations of the
super three-form potential $B$. Superdiffeomorphisms are
characterized by super Killing vectors $\xi^A$, while gauge
transformations are parameterized by a super two-form $\L_{MN}$. The
commutator of two diffeomorphisms, $\xi^A$ and ${\xi'}^A$, simply leads
to a third diffeomorphism ${\xi''}^A$ given by the super-Lie bracket of
$\xi^A$ and ${\xi'}^A$. It is easy to show that the commutator of a
superdiffeomorphism $\xi^A$ with a gauge transformation $\L$ is a second gauge
transformation with a parameter
\begin{equation}
   \L' = \iota_\xi d \L + d \iota_\xi \L
\end{equation}
The commutator of two gauge transformations is, of course, zero. It is
always possible to define transformations as combinations of
diffeomorphisms and gauge transformations. The supersymmetry
transformations in WZ gauge defined in section three are such an
example. Such transformations will have a more complicated algebra,
but this is an artifact of their definition. The underlying algebra is
always superdiffeomorphisms together with gauge transformations. 

We now want to consider a specific solution of the supergravity
theory, namely flat space, and ask which of the local symmetries leave
the solution invariant. From the bosonic action \eqref{eq:4-5} we see
that flat space is a solution of three-form supergravity with
$F_{mnpq}=0$. The constraints on the other components of
$\Si$ \eqref{sigma_constraints} imply that the super four-form
completely vanishes except for the components
\begin{equation}
   \Sigma_{\d\g ba} = 
      - \frac{2}{3}\left(\s_{ba}\e\right)_{\d\g}  \quad
   {\Sigma^{\dd\gd}}_{ba} = 
      - \frac{2}{3} \left( \bar{\s}_{ba}\e\right)^{\dd\gd}
\label{flatSigma} 
\end{equation}
Thus, the flat-space solution in fact corresponds to a constant super
four-form background. 

If we ignore the four-form, we know that the symmetries of flat
superspace are simply ordinary translations and Lorentz rotations, and
supertranslations, which are just the flat-space supersymmetry
transformations. Together these form the usual
super-Poincar\'e algebra. From the form of the flat-space field strength
\eqref{flatSigma} it is clear that these symmetries also leave $\Si$
invariant. We are also free to make local gauge transformations. Thus,
requiring $\Si$ to be invariant, the symmetries of flat space are
simply the usual unextended super-Poincar\'e transformations together
with local gauge transformations. 

We now ask a more restrictive question: what is the algebra
of symmetries of flat space which leave the super three-form potential
itself invariant? We will argue shortly, that this is the algebra
relevant to matter which carries form-field charge. 
Given that $\Si$ is not zero, clearly the potential $B$ is also
non-zero. We are in flat space, so can use the discussion of
section two to find an expression for $B$. The constraints on
$\Si$ imply that the components of the chiral superfield $Y$ are set to
$Y=-4/3$ and $\eta_\a=H=0$ (see \eqref{eq:3-39} and
\eqref{eq:3-40}). Since here $F_{mnpq}=0$, we can also choose a gauge
where $C_{mnp}=0$. Comparing with equation \eqref{eq:2-30}, we see
that there is a gauge where $\Omega$ takes the form
\begin{equation}
   \Omega = -\frac{1}{12} \left( \theta^2 + \bar{\theta}^2 \right)
\end{equation}
Equations \eqref{eq:2-18}--\eqref{eq:2-20} imply that all
the components of $B$ vanish except for 
\begin{equation}
\begin{gathered}
   B_{\a\ \ a}^{\ \;\bd} = 
      \frac{i}{6} \left(\theta^2+\bar{\theta}^2\right) 
         \left(\s_a\e\right)_{\a}^{\ \;\bd} \\
   B_{\a ba} = - \frac{1}{3} {\left(\s_{ba}\right)_\a}^\b \th_\b \qquad
   {B^\ad}_{ba} = - \frac{1}{3} {\left(\sb_{ba}\right)^\ad}_\bd \bar{\th}^\bd 
\end{gathered}
\label{flatB}
\end{equation}
That $B$ does not vanish is to be expected, since we know that even in
flat space, $\kappa$-symmetry requires a non-zero Wess-Zumino term in
the Green-Schwarz membrane action. This non-trivial term is none other
than the coupling of the membrane to a background $B$ field. 

We note first that, in this gauge, $B$ is invariant under normal
translations. However it is not invariant under supertranslations. 
This is easily seen from \eqref{flatB}, which clearly changes 
under $\th^\a\rightarrow\th^\a+\xi^\a$. However we know that
the field strength $\Si=dB$ is invariant under supertranslations. This
implies that the variation of $B$ must simply be a gauge
transformation. That is, it is possible to write it in the form
\begin{equation}
   \d_\xi B = d \L_\xi
\label{Bvar}
\end{equation}
where $\L_\xi$ is a particular two-form dependent on $\xi^\a$. Clearly
if we want to leave $B$ invariant we must modify the supertranslation
transformation. We define a new variation which combines the
supertranslation with a gauge transformation,
\begin{equation}
   \d^{\text{new}}_\xi B = \d_\xi B + \d_{\L_\xi} B
\label{newsusy}
\end{equation}
where $\d_{\L_\xi}B=-d\L_\xi$, so that we have $\d^{\text{new}}_\xi
B=0$ as required. The only remaining transformations which preserve
$B$ are a trivial subset of the general gauge transformations, for
which $\d_\L B=d\L=0$. 

What is the algebra of these symmetries? The easiest way to find out 
is to use the supersymmetry transformations given in section two. We
recall that the variation \eqref{eq:2-6} did not correspond to simple
supertranslation but included a gauge transformation in order to
preserve the WZ gauge. In fact, it is exactly the modified transformation
we defined above \eqref{newsusy} in order to leave $B$ invariant. 
To see this, recall that the three-form in
the supergravity theory was constrained so that $Y=-4/3$ and
$\eta_\a=H=0$ (see \eqref{eq:3-39} and \eqref{eq:3-40}). Thus in WZ
gauge the only surviving component of $B$ is $C_{mnp}$. Its variation
is given by \eqref{eq:2-6}, which vanishes since $\eta_\a=0$. Thus the
WZ-gauge supersymmetry variations of section two leave $B$ invariant,
and correspond exactly to the modified supertranslation
\eqref{newsusy}. The commutator of two of the modified
supertranslations can be taken directly from \eqref{eq:2-36},
\begin{equation}
   \left(\d \d'-\d' \d \right) C_{mnp} = 
       2\eta^{q}\partial_{q}C_{mnp} + 3\p_{[m}\L_{np]}
\end{equation}
where we have dropped the label `new' and 
\begin{equation}
   \Lambda_{mn} = -2C_{mnp}\eta^{p} 
       - \frac{1}{2} \left( \eta^{+}_{pq}Y_{1} + \eta^{-}_{pq}Y_{2} \right)
\end{equation}
with
\begin{equation}
\begin{aligned}
   \eta^{m} &= -i \left( \xi\s^m\bar{\xi}' - \xi'\s^m\bar{\xi} \right) \\
   \eta^{+}_{pq} &= \xi\s_{pq}\xi' + \bar{\xi}\bar{\s}_{pq}\bar{\xi}' \\
   \eta^{-}_{pq} &= -i \left( \xi\s_{pq}\xi' 
       - \bar{\xi}\bar{\s}_{pq}\bar{\xi}' \right) \\
\end{aligned}
\end{equation}
Recalling that $Y=Y_1+iY_2=-4/3$ and that we have chosen the gauge
$C_{mnp}=0$, we find the gauge transformation parameter simplifies to
$\L_{mn}=\frac{2}{3}\eta^+_{mn}$, so that 
\begin{equation}
   \left(\delta\delta'-\delta'\delta\right) C_{npq} = 
      2\eta^r \p_r C_{npq} + 2\partial_{[n} \eta^+_{pq]}  
\end{equation}
Thus the commutator of two of the modified
supertranslations closes into an ordinary translation, parameterized by
$\eta^m$, together with a gauge transformation, parameterized by
$\eta^+_{mn}$. This can now be compared with the extended algebra
\eqref{Zalgebra}. Given the form of $\eta^r$ and $\eta^+_{pq}$ we see
that, up to a normalization of the charge $Z_{mn}$, the two are identical. 
The extended superalgebra is none other
than the algebra of the symmetries of flat space which leave $B$
invariant. 

Note that we derived the extended algebra in a particular
gauge. However, we would have in fact obtained the same form whatever 
gauge we had started in. The only modification would be in the
definition of the modified translation and supertranslation
symmetries. Suppose, for instance, we chose a gauge where $C_{mnp}$
was non-zero. Under an ordinary translation
\begin{equation}
   \d_\eta C_{mnp} = \eta^r \p_r{C_{mnp}} \neq 0
\end{equation}
Thus we must redefine a modified translation symmetry which includes a
gauge transformation, such that
\begin{equation}
   \d^{\text{new}}_\eta C_{mnp} = \eta^r\p_r{C_{mnp}} 
           - 6\p_{[m}\left(\eta^r C_{np]r}\right) 
       = \eta^r\Si_{rmnp} = 0
\end{equation}
and so $C_{mnp}$ is invariant. Using this modified symmetry, one then finds
the same extended algebra. It turns out that there is a
gauge-invariant property of $\Si$ which is leading to the extension of
the algebra \cite{AT}. The point is that while the field strength
$\Si$ was invariant under the super-Poincar\'e group, it was
impossible to find a gauge where the potential $B$ was also
invariant. Formally, $\Si$ is an element of the fourth
Chevalley-Eilenberg cohomology class of the super-translation group,
where flat space is considered as the group manifold. The central
extension of the algebra is then the analog of the usual Abelian
extension of a Lie algebra with a non-trivial second cohomology class
(\cite{AdeA} and references therein). 

The question remains as to why the algebra of the symmetries of
flat-space which leave $B$ invariant should correspond to the
topologically-extended algebra which describes membrane states. We
start by noting that a membrane is the archetypical object which
carries three-form charge. Whether it is fundamental or a
solitonic solution, a membrane in a flat background will be
effectively described by a Green-Schwarz action. It will couple to the
three-form via the Wess-Zumino term 
\begin{equation}
   S_{\text{WZ}} = 
       \int{} d^3\s\; \sqrt{-\g} \e^{ijk} {E_i}^A{E_j}^B{E_k}^C B_{CBA}
\label{WZaction}
\end{equation}
Under what symmetries is this term invariant? The projected vielbeins
${E_i}^A=\p_i{Z}^M{E_M}^A$ are invariant under the usual translations
and supertranslations by construction, and also trivially under any gauge
transformations. But the three-form $B$ is only invariant under the
set of modified transformations we described above. Thus the symmetry
algebra of the Green-Schwarz action is exactly the algebra of the
symmetries which leave $B$ invariant. 

This also allows us to
make the connection to the calculation of the extended algebra given in
\cite{AGITb}. There it was noted that the WZ term was invariant under
supertranslations only up to a total divergence. In the present
context we can understand this result as a reflection of
\eqref{Bvar}. Namely the variation of $B$ under a supertranslation is
just a gauge transformation. It is clear from the form of \eqref{WZaction}
that a gauge transformation in $B$ leads to a total divergence. The
authors of \cite{AGITb} defined a conserved current by subtracting the
total divergence term from the conventional supertranslation current. 
But this is exactly the current one would define for a transformation
which included a compensating gauge transformation together with the
supertranslation. That is, it is just the current corresponding to the
modified supertranslation \eqref{newsusy}. 

We recall from the solution of section four, that the actual extended
membrane does not live in asymptotically flat space in four
dimensions. Since there is no fall off in the gauge field away from
the membrane, the asymptotic space is actually anti-de Sitter. Given
this, one might ask how the supersymmetry algebra is extended in a
anti-de Sitter background. Calculating the algebra from the solution
itself is problematic since it is not clear how to define the
asymptotic charges. The patching of two pieces of anti-de Sitter space
means that there is no clear sense in which the membrane is a
perturbation about an anti-de Sitter background. However, we can ask
what symmetries leave the Wess-Zumino term in the membrane action
invariant. That is, find the algebra of symmetries that leave $B$
invariant, as we did in the flat case. 

The fermionic part of the anti-de Sitter superalgebra is given by
\begin{equation}
\begin{gathered}
   \big\{Q_\a,\Qb_\ad\big\} = 2 {\s^m}_{\a\ad} J_{m4} \\
   \big\{Q_\a,Q_\b\big\} = 2i \big(\s^{mn}\big)_{\a\b} J_{mn} \qquad
   \big\{\Qb_\ad,\Qb_\bd\big\} = 2i \big(\sb^{mn}\big)_{\ad\bd} J_{mn} \\
   \big[J^{m4},Q_\a\big] = \frac{i}{4}{\s^m}_{\a\ad} \Qb^\ad \quad
   \big[J^{m4},\Qb_\ad\big] = \frac{i}{4}{\sb^m}_{\ \ \ad\a} Q^\a \\
   \big[J^{mn},Q_\a\big] = 
       -\frac{1}{2}{\big(\s^{mn}\big)_\a}^\b Q_\b \quad
   \big[J^{mn},\Qb^\ad\big] = 
       -\frac{1}{2}{\big(\sb^{mn}\big)^\ad}_\bd \Qb^\bd
\end{gathered}
\label{adSSUSY}
\end{equation}
where $J^{mn}$, $J^{m4}$ are the generators of the $SO(3,2)$ group of
isometries of anti-de Sitter space. As in the previous example we can
use the supersymmetry variations from section three to calculate the
algebra, since these leave the three-form $B$ invariant. The full
calculation requires explicit forms for the Killing vectors and spinors
in anti-de Sitter space and we will not give the details here. The
result is simply that there is no extension of the algebra. A membrane state
in anti-de Sitter space is just a representation of the usual anti-de
Sitter supersymmetry algebra \eqref{adSSUSY}. In fact, oscillations of the 
flat membrane have been identified with the singleton representation 
\cite{single}. While it can carry three-form
charge, this charge does not enter the supersymmetry
algebra. Algebraically, this is because there is no corresponding
central extension of the super-anti-de Sitter algebra possible. The
extension must be central because gauge transformations commute, and
no such extension by a two-form $Z_{mn}$ exists.



\end{document}